\documentclass[aps,prl,twocolumn,superscriptaddress,showpacs]{revtex4-2}
\usepackage{graphicx}
\usepackage{dcolumn}
\usepackage{bm}
\usepackage{subfigure}
\usepackage{amsmath}
\usepackage{amssymb}
\usepackage{hyperref}
\usepackage{siunitx}

\newcommand{\bk}{{\boldsymbol k}}

\newcommand{\ua}{\uparrow}
\newcommand{\da}{\downarrow}

\newcommand{\SRO}{Sr$_2$RuO$_4$} 

\usepackage{times}

\begin{document}
\title{Quantum-geometry-induced anomalous Hall effect in nonunitary superconductors and application to \SRO}
	\author{Jia-Long Zhang} 
	\thanks{These two authors contributed equally to this work.}
	\address{Department of Physics, The Hong Kong University of Science and Technology, Hong Kong SAR, China}
	\address{Shenzhen Institute for Quantum Science and Engineering, Southern University of Science and Technology, Shenzhen 518055, Guangdong, China}
	\author{Weipeng Chen}
	\thanks{These two authors contributed equally to this work.}
	\address{Shenzhen Institute for Quantum Science and Engineering, Southern University of Science and Technology, Shenzhen 518055, Guangdong, China}
	\address{International Quantum Academy, Shenzhen 518048, China}
	\address{Guangdong Provincial Key Laboratory of Quantum Science and Engineering, Southern University of Science and Technology, Shenzhen 518055, China}
	\author{Hao-Tian Liu}
	\address{Shenzhen Institute for Quantum Science and Engineering, Southern University of Science and Technology, Shenzhen 518055, Guangdong, China}
	\address{International Quantum Academy, Shenzhen 518048, China}
	\address{Guangdong Provincial Key Laboratory of Quantum Science and Engineering, Southern University of Science and Technology, Shenzhen 518055, China}
	\author{Yu Li}
     \address{Institute of Quantum Materials and Physics, Henan Academy of Sciences, Zhengzhou, Henan 450046, China}
     \address{Kavli Institute for Theoretical Sciences, University of Chinese Academy of Sciences, Beijing 100190, China}
\author{Zhiqiang Wang}
\address{Department of Physics and James Franck Institute, University of Chicago, Chicago, Illinois 60637, USA}
    \author{Wen Huang}
	\email{huangw3@sustech.edu.cn}
	\address{Shenzhen Institute for Quantum Science and Engineering, Southern University of Science and Technology, Shenzhen 518055, Guangdong, China}
	\address{International Quantum Academy, Shenzhen 518048, China}
	\address{Guangdong Provincial Key Laboratory of Quantum Science and Engineering, Southern University of Science and Technology, Shenzhen 518055, China}

\date{\today}

\begin{abstract}
The polar Kerr effect and the closely related anomalous charge Hall effect are among the most distinguishing signatures of the superconducting state in \SRO, as well as in several other compounds. These effects are often thought to be derived from chiral superconducting pairing, and different mechanisms have been invoked for the explanation. However, the intrinsic mechanisms proposed previously often involve unrealistically strong interband Cooper pairing. In this study we show that, even without interband pairing, nonunitary superconducting states can support intrinsic anomalous charge Hall effect, thanks to the quantum geometric properties of the Bloch electrons. The key here is to have a normal-state spin Hall effect, for which a nonzero spin-orbit coupling is essential. A finite charge Hall effect then naturally arises at the onset of a spin-polarized nonunitary superconducting pairing. It depends on both the spin polarization and the normal-state electron Berry curvature, the latter of which is the imaginary part of the quantum geometric tensor of the Bloch states. Applying our results to the weakly-paired \SRO~we conclude that, if the reported Kerr effect is of intrinsic origin, the superconducting state is most likely nonunitary and has odd-parity. Our theory may be generalized to other superconductors that exhibit polar Kerr effect. 
\end{abstract}

\maketitle
{\bf Introduction --} Cooper pairs in chiral superconductors carry a nonzero and quantized orbital angular momentum~\cite{Kallin:16}. A condensate of such time-reversal-symmetry-breaking (TRSB) Cooper pairs may support spontaneous Hall-like response even in the absence of external magnetic field, {\it i.e.} an anomalous Hall effect~\cite{Xiao:10,Nagaosa:10}. This Hall effect at optical frequency is directly related to the polar Kerr effect reported in a number of much-debated chiral superconductor candidates, including \SRO~\cite{Xia:06}, UPt$_3$~\cite{Schemm2014}, and URu$_2$Si$_2$~\cite{Schemm2015}. However, understanding the Hall effect in chiral superconductors is much more challenging than in quantum Hall insulators. General Galilean invariance principle~\cite{Read:00} dictates that the Hall conductivity should vanish in a clean single-band chiral superconductor, because the center-of-mass (COM) motion of a Cooper pair subject to external electric field is oblivious to the relative motion between the two paired electrons (and is, therefore, independent of the chiral nature of the pairing). In a semiclassical perspective, this invariance is related to the vanishing of the anomalous velocity~\cite{Karplus:54,Sundaram:99} associated with Bogoliubov quasiparticles that travel under the influence of the electric field.

One way to break the Galilean invariance and hence to entangle the relative and COM motion is by breaking translation symmetry~\cite{Goryo:08,Lutchyn:09,Konig:17,LiYu:20}, which can be achieved by extrinsic disorder. The invariance may also be broken, in an intrinsic manner, in clean multiband superconductors~\cite{Taylor:12,Wysokinski:12,Komendova:17,WangZQ:17,Brydon:19,ZhangJL:20}. Here, the requirement for sizable interband Cooper pairing is implicit in the analyses~\cite{Taylor:12,Wysokinski:12,Komendova:17,WangZQ:17,Brydon:19,ZhangJL:20}. Owing to the different group velocities on different bands, a generic interband pair of electrons at opposite momenta typically carries a nonvanishing COM momentum. Note that most of the previous multiband studies were performed in the orbital-basis language. Hence, when translated into the band-basis description, the pertinent models almost by default encapsulate comparable intraband and interband pairings. This has been met with skepticism~\cite{Mineev:12}, as the weak superconductivity in the above-mentioned superconductors is not expected to develop interband pairing of the size needed to explain the observed Kerr rotation angle. 

Another necessary ingredient for the intrinsic anomalous Hall effect in chiral superconductors is the interband velocity~\cite{Taylor:12,Denys:21}, {\it i.e.}~the off-diagonal elements of the velocity matrix in band basis. The underlying physics, however, has not been fully elucidated. In fact, this quantity has its origin in the quantum geometric properties of the Bloch bands~\cite{Peotta:15,Torma:21,AhnJ:22}. Specifically, the $\mu$-th component of the interband velocity is given by~\cite{Liang:17,Chen:21}
\begin{equation}
V^{ij}_{\mu\bk} = (\epsilon_{i\bk} -\epsilon_{j\bk}) \langle  \partial_\mu \psi_{i\bk}| \psi_{j\bk}\rangle\,,~~~(\epsilon_{i\bk} \neq \epsilon_{j\bk})
\label{eq:Vij}
\end{equation}
where $\partial_{\mu}\equiv \partial_{\bk_\mu}$ and $\{i,j\}$ label the normal-state electron energy bands. $\epsilon_{i\bk}$ and $|\psi_{i\bk}\rangle$ denote the respective energy dispersion and eigenvector of the $i$-th band. The object $i\langle \partial_{\mu}\psi_{i\bk}| \psi_{j\bk}\rangle$ defines a non-Abelian Berry connection between the two Bloch states, which is closely related to the definition of the quantum geometric tensor~\cite{Provost:80},
\begin{equation}
g^i_{\mu\nu,\bk} = \langle \partial_{\mu}\psi_{i\bk}|\partial_{\nu}\psi_{i\bk}\rangle-  \langle \partial_{\mu}\psi_{i\bk}|\psi_{i\bk}\rangle  \langle\psi_{i\bk}|\partial_{\nu}\psi_{i\bk}\rangle \,.
\label{eq:Gtensor}
\end{equation}
Put simply, the interband velocity describes a concerted motion of electrons from different bands, {\it i.e.}~their charge transport is not independent of each other, even though they belong to distinct energy eigenstates. Such a quantum connection therefore provides another intrinsic route, besides the interband pairing, to entangle the relative and COM motion of a Cooper pair, and hence to break the aforementioned Galilean invariance. 

A question then follows: {\it without the disputable interband pairing, is the finite interband velocity alone sufficient to support an intrinsic charge Hall response in chiral or other TRSB superconductors?} In this study, we answer this question in the affirmative. Two ingredients are key to our proposal. One is spin-orbit coupling (SOC), which results in a finite spin Hall effect in the normal state. The other is nonunitary odd-parity pairing, which causes imbalanced spin occupancy. Taken together, these two ingredients naturally lead to finite anomalous charge Hall effect and consequently polar Kerr effect. The charge Hall conductivity is related to the superconducting spin polarization, as well as the normal-state Berry curvature of the Bloch electrons -- which is given by the imaginary part of the geometric tensor in Eq.~\eqref{eq:Gtensor} and which vanishes in the absence of SOC. 

In fact, a previous study~\cite{Denys:21} has already pointed out that anomalous charge Hall response can arise in a multiband superconductor with an intraband-only but nonunitary pairing. However, neither the requirement for finite SOC nor the relation to quantum geometry ({\it i.e.}~the aforementioned Bloch Berry curvature) was made explicit. Our work fills in these gaps and thereby provides a deeper and more intuitive understanding of the mechanism. 

Below, we focus on the compound \SRO~\cite{Xia:06}~to illustrate the physics, although the discussions can be generalized to other superconductors as well. While it has been studied for about 30 years, the pairing symmetry of \SRO~is still under vigorous debate~\cite{Maeno2001,Maeno2003,Kallin2012, Maeno2012, Mac2017, Huang2021CPB, Leggett2021}. A number of early observations point to chiral $p$-wave pairing~\cite{Luke1998, Ishida1998,Duffy2000, Liu2004,Xia:06}; however, this has been challenged by recent experimental advances~\cite{Hicks:14,Pus2019,Chronister2021,exp1,exp3,ultrasound1,ultrasound2}, notably the NMR measurements which have completely reshaped our research landscape. Various contending candidate pairing symmetries have been proposed. However, none of them can coherently interpret all key observations. Specific to the Kerr effect, it is only compatible with states that break all vertical mirror plane symmetries~\cite{Cho:16}. Thus chiral states such as $p+ip$ or $d+id$ and nonunitary helical p-wave states~\cite{Huang2021prr} emerge as prominent candidates~\cite{footnote}. The present study places further constraints on the pairing symmetry of \SRO~and several other compounds, if the reported Kerr effect~\cite{Xia:06,Schemm2014,Schemm2015} is dominated by intrinsic contribution ({\it i.e.}, not by extrinsic disorder scattering effects) and if interband pairing is irrelevant in these materials.

{\bf Nonunitary p-wave pairings in \SRO --} In accordance with the $D_{4h}$ crystallographic symmetry of \SRO, two types of elemental nonunitary states have been discussed in the literature~\cite{Huang2021CPB,Huang2018prl,Huang2021prr}. 
One is formed by complex mixtures of distinct helical $p$-wave channels~\cite{Huang2021prr}, {\it i.e.}~$A_{1u}+iA_{2u}$ and $B_{1u}+iB_{2u}$, where $\{A_{1u}, A_{2u}, B_{1u}, B_{2u} \}$ represent the four one-dimensional and odd-parity irreducible representations ({\it irreps}) of the $D_{4h}$ group. These states can be viewed as a combination of $p+ip$ and $p-ip$ pairings in two respective spin sectors, with unequal pairing amplitudes. As an example, the gap function of the $A_{1u}+iA_{2u}$ state can be written as
\begin{equation}
\hat{\Delta}_{A_{u}} =\begin{pmatrix} \Delta_{\ua\ua}    & \Delta_{\ua\da} \\
                                                   \Delta_{\da\ua} & \Delta_{\da\da} \end{pmatrix} = \begin{pmatrix} \Delta(-k_x+ik_y)    & 0 \\
                                                   0& \Delta^\prime(k_x+ik_y) \end{pmatrix} \,,
\label{eq:mixedHelical}
\end{equation}
where the amplitude of the spin-up and spin-down pairings are $\Delta = \Delta_{A_{1}}+\Delta_{A_{2}}$, $\Delta^{\prime} = \Delta_{A_{1}}-\Delta_{A_{2}}$. Here, $\Delta_{A_{1}}$ and $\Delta_{A_{2}}$ denote the amplitude of the $A_{1u}$ and $A_{2u}$ components, respectively. The nonunitary nature is evident because $|\Delta_{\ua\ua}| \neq |\Delta_{\da\da}|$. Note that, in the presence of SOC, the spins here should be understood as pseudospins.  

The other type of nonunitary pairing is three-dimensional (3D), has a chiral $p$-wave symmetry, and belongs to the $E_u$ {\it irrep}. The chiral $p$-wave traditionally discussed in the context of \SRO~has the in-plane pairing form of $(k_x+ik_y)\hat{z}$. Here, $\hat{z}$ represents the component of $\bf{d}$-vector which depicts the spin configuration of a spin-triplet pairing. This form characterizes a Cooper pair with orbital angular momentum $L_z=\hbar$ and spin angular momentum $S_z=0$. When spin rotation symmetry is broken by SOC, $L_z$ and $S_z$ are no longer good quantum numbers. In this case, the above pairing shall in general mix with another component that features $L_z=0$ and $S_z= \hbar$ since both have the same $J_z=L_z+S_z=\hbar$. The latter turns out to be an out-of-plane $k_z$-like pairing $k_z(\hat{x}+i\hat{y})$. The resultant pairing acquires a 3D structure with the following general form~\cite{Huang2018prl},
\begin{equation}
\hat{\Delta}_{E_u} = \begin{pmatrix} 2\Delta_\perp k_z    & \Delta_{\parallel}(k_x+ik_y) \\
                                                   \Delta_{\parallel}(k_x+ik_y) & 0 \end{pmatrix} \,,
\label{eq:3DchiralPwave}
\end{equation}
where $\Delta_{\parallel (\perp)}$ denotes the amplitude of the in-plane (out-of-plane) pairings.  

{\bf Normal-state spin Hall effect --} Much of the physics can be captured by a two-orbital model consisting of Ru $d_{xz}$ and $d_{yz}$ orbitals residing on a square lattice. The corresponding tight-binding Hamiltonian can be written as, 
\begin{equation}
H_\bk = \epsilon_{\bk}+\tilde{t}_{\bk} \sigma_z + \lambda_{\bk}\sigma_x + \eta_{\bk} \sigma_y s_z \,,
\label{eq:2bandTB}
\end{equation}
where the Pauli matrices $\sigma_i$ and $s_i$ operate respectively on the orbital and spin subspace, $\epsilon_{\bk} = t(\cos k_x + \cos k_y) -\mu$, $\tilde{t}_{\bk}=\tilde{t}(\cos k_x -\cos k_y)$, $\lambda_{\bk}=\lambda\sin k_x\sin k_y$, and $\eta_{\bk}= \eta_0 $. Here, $\{t,\tilde{t}\}$ describe the hopping integrals, $\mu$ is the chemical potential, $\lambda$ quantifies the orbital hybridization, and $\eta_{0}$ denotes the leading order of SOC. The above Hamiltonian preserves time-reversal and inversion symmetries, and it gives two sets of Kramers-degenerate Bloch bands, $H_\bk|\psi_{i\bk s}\rangle = \epsilon_{i\bk} |\psi_{i\bk s}\rangle$ with $i=\{1,2\}$ and $s=\{\uparrow,\downarrow\}$. 
Note that, since the present SOC term preserves the $U(1)$ spin rotation symmetry about the $z$-axis, a convenient basis is available where $s$ represent real spins quantized in the $z$-direction. More generally, higher order SOC terms, such as $\eta^\prime\sigma_y \sin k_z(\sin k_x  s_x + \sin k_y s_y)$, are allowed by symmetry, and they break this remaining $U(1)$ symmetry. Nonetheless, one can always choose a smooth gauge under which the orientation of the pseudospins evolves continuously in momentum space. In this case, the matrix element of the velocity operator in the band basis can be expressed as,
\begin{equation}
V_{\mu\bk s}^{ij} = \delta_{ij}\partial_\mu \epsilon_{i\bk} + (\epsilon_{i\bk}-\epsilon_{j\bk})\langle \partial_\mu \psi_{i\bk s}|\psi_{j\bk s}\rangle \,,
\label{eq:velocity}
\end{equation}
where $\delta_{ij}$ is the Kronecker delta function. The second term above is the interband velocity. By time-reversal and hermiticity, one has $V^{ij}_{\mu\bk s} = -(V^{ij}_{\mu\bar{\bk}\bar{s}})^\ast = (V^{ji}_{\mu\bk s})^\ast$ with $\bar{s}=-s$ and $\bar{\bk}=-\bk$. 

\begin{figure}
\includegraphics[width=8.5cm]{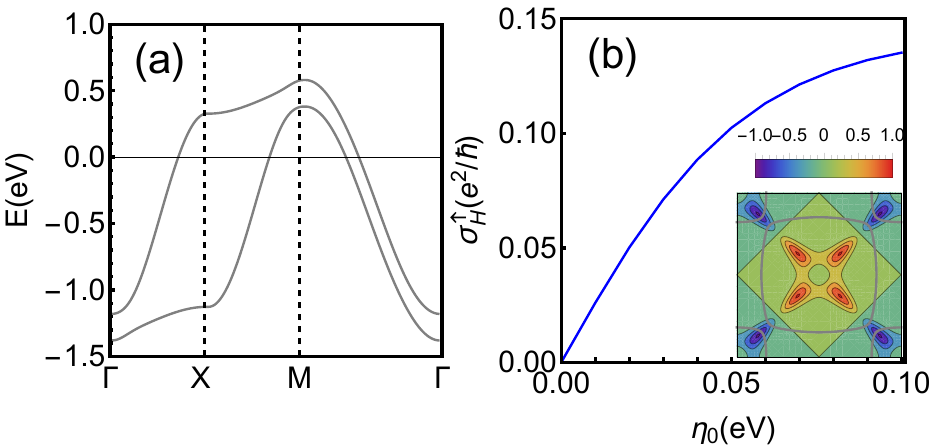}
\caption{(a) Band structure of the model in Eq.~\eqref{eq:2bandTB} with parameters $(t,\tilde{t},\lambda,\eta_0,\mu)=(-0.44, -0.36, 0.08, 0.1, 0.4)$ eV. There are two energy bands, each of which is two-fold Kramers degenerate. (b) The corresponding spin-up and zero-temperature charge Hall conductivity $\sigma_{H}^{\uparrow}$, which is directly related to the spin Hall conductivity $\sigma_{H}^{\text{spin}}$ in Eq.~\eqref{eq:spinHall}. The inset shows the momentum-space distribution of the spin-up Berry curvature of the upper band in (a). The thick grey curves in the inset depict the Fermi surfaces.}
\label{fig:Berry}
\end{figure}

One can further define the pseudospin-dependent quantum geometric tensor for band-$i$, $g^{i}_{\mu\nu,\bk s}$, as in Eq.~\eqref{eq:Gtensor}. The real part of this tensor, {\it i.e.} the quantum metric~\cite{Provost:80,Torma:21}, is nonvanishing as long as either the orbital hybridization $\lambda$ is finite or a momentum-dependent SOC is present. The imaginary part of the tensor is the Berry curvature,
\begin{eqnarray}
\mathcal{B}^{i}_{\mu\nu,\bk s} &=& -2\text{Im}[g^{i}_{\mu\nu,\bk s}] = i\sum_{j \neq i} \frac{V^{ij}_{\mu\bk s} V^{ji}_{\nu\bk s} - V^{ij}_{\nu\bk s} V^{ji}_{\mu\bk s}}{(\epsilon_{i\bk}-\epsilon_{j\bk})^2} \nonumber \\
&=& i \big[  \langle\partial_\mu\psi_{i\bk s}|\partial_\nu \psi_{i\bk s}\rangle  -   \langle\partial_\nu\psi_{i\bk s}|\partial_\mu \psi_{i\bk s}\rangle \big].
\label{eq:BerryCurvature}
\end{eqnarray}
Note that the Berry curvature is finite only when both orbital hybridization and SOC are present. Figure \ref{fig:Berry} (c) shows the spin-up Berry curvature obtained from Eq.~\eqref{eq:2bandTB} for one of the bands~\cite{footnote2}. The spin-down Berry curvature of the corresponding Kramers degenerate band is related by time-reversal and is thus opposite in sign. Hence the normal state of the present system exhibits spin Hall but no charge Hall effect, as noted in an earlier work~\cite{Imai:12}. The spin Hall conductivity can be qualitatively described by~\cite{Haldane:04,Hsu2011}
\begin{equation}
\sigma^\text{spin}_H = \left(\frac{e}{2}\right) \frac{2}{N} \sum_{i=1,2} \sum_\bk ~\mathcal{B}^{i}_{x y,\bk\ua} f(\epsilon_{i\bk}) \,,
\label{eq:spinHall}
\end{equation}
where $N$ denotes the size of the system, a prefactor 2 accounts for the two spin contributions, $e/2$ is added to give the correct dimension of spin Hall conductivity~\cite{Hsu2011}, and $f(\epsilon)$ is the Fermi distribution function. The $\bk$-summation runs over the first Brillouin zone. At small SOC, $\sigma^\text{spin}_H$ increases linearly with $\eta_0$, as can be seen in Fig.~\ref{fig:Berry} (b).  

{\bf Intrinsic charge Hall effect --} At the onset of a spin-polarized nonunitary pairing, the charge Hall conductivity of the two spin species no longer cancel exactly, leading naturally to an intrinsic anomalous charge Hall response. Within linear response theory, the charge Hall conductivity of the superconducting state at optical frequency $\omega$ is defined as~\cite{Taylor:12,Wysokinski:12}, 
\begin{equation}
\sigma_H(\omega) = \frac{i}{2\omega}[\pi_{xy}(\omega+ i 0^+ ) - \pi_{yx}(\omega + i 0^+)] \,,
\end{equation}
in which $\pi_{xy}$ is the transverse current-current correlator, which in the Matsubara-frequency representation is given by $\pi_{xy}(i\nu_m)=(e^2/2 \hbar^2) \sum_{\bk} (k_BT) \sum_{\omega_n}\text{Tr}[\mathcal{V}_{x\bk} G(\bk,i\omega_n) \mathcal{V}_{y\bk} G(\bk,i\omega_n+i\nu_m)]$. Here $\omega_n$ and $\nu_m$ are the fermionic and bosonic Matsubara frequencies, respectively; $G(\bk,i\omega_n) = (i\omega_n - H^\text{BdG}_{\bk})^{-1}$ denotes the Gor'kov Green's function of the corresponding Bogoliubov de-Gennes (BdG) Hamiltonian $H^\text{BdG}_{\bk}$. In the band basis with the Nambu spinor $(c_{1\bk\ua},c_{1\bk\da},c_{2\bk\ua},c_{2\bk\da},c^\dagger_{1\bar{\bk}\ua},c^\dagger_{1\bar{\bk}\da},c^\dagger_{2\bar{\bk}\ua},c^\dagger_{2\bar{\bk}\da})^T$, the corresponding (charge-current) velocity operator matrix is given by
\begin{equation}
\mathcal{V}_{\mu\bk} = \begin{pmatrix} V_{\mu\bk}  &   \nonumber \\
                                                    &  -(V_{\mu\bar{\bk}})^{*}
                             \end{pmatrix} \,,
\end{equation}
where the matrix elements of $V_{\mu \bk}$ are given in Eq.~\eqref{eq:velocity}. To expose the virtual optical transition processes responsible for the Hall response, we use the spectral representation of the Green's function to derive the Hall conductivity and obtain
\begin{eqnarray}
\sigma_{H}(\omega)=\frac{i e^2}{4N\hbar \omega}\sum_{a,b,\bk}\frac{f(E_{a\bk})-f(E_{b\bk})}{ \hbar \omega+i 0^{+}+E_{a\bk}-E_{b\bk} } Q^{ab}_{xy,\bk}\,, \nonumber \\
\label{eq:AHE} 
\end{eqnarray}
with
\begin{equation}
Q^{ab}_{xy,\bk}= \langle b\bk|\mathcal{V}_{x\bk}|a\bk\rangle\langle a\bk|\mathcal{V}_{y\bk}|b\bk\rangle-( x \leftrightarrow y) \,.
\end{equation}
Here, $|a\bk\rangle$ designates the $a$-th Bogoliubov quasiparticle solution, \textit{i.e.}~$H^\text{BdG}_{\bk} |a\bk\rangle = E_{a\bk}|a\bk\rangle$ where $E_{a\bk}$ contains both positive and negative branches.

\begin{figure}
\includegraphics[width=5cm]{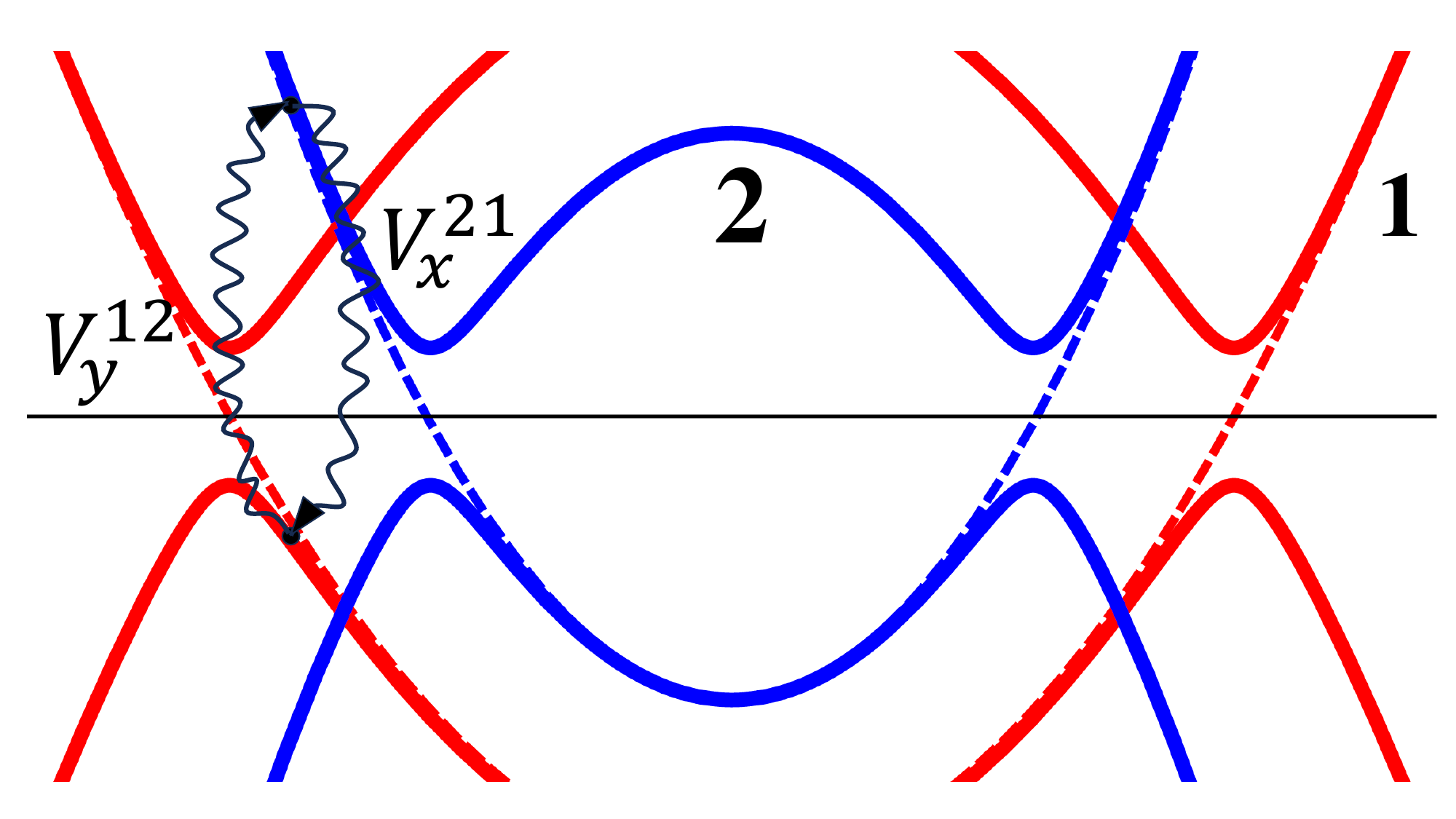}
\caption{Sketch of representative virtual optical transitions responsible for the Hall response in one pseudospin sector of the nonunitary helical $p$-wave state. The red and blue solid curves represent the Bogoliubov spectra of the two superconducting bands (labelled by $1$ and $2$) when interband pairing is absent. Their corresponding normal state electronic dispersions are plotted in dashed curves with respective colors.  }
\label{fig:transitions}
\end{figure}

In single-band superconductors without SOC, the transition matrix element $\langle a\bk|\mathcal{V}_{\mu\bk}|b\bk\rangle$ vanishes for $a\neq b$, because in that case the velocity operator commutes with the BdG Hamiltonian and, as a result, $\pi_{xy}(\omega) \equiv \pi_{yx}(\omega)$. In multiband superconductors without interband Cooper pairing, the Bogoliubov quasiparticle solutions associated with different Bloch bands are fully decoupled. However, the velocity operators in general do not commute with the Hamiltonian. Specifically, interband velocity may couple quasiparticles from different bands, thereby enabling interband transitions. In other words, the aforementioned transition matrix element may not vanish if the two quasiparticle states originate from different Bloch bands. A pair of interband transition processes induced by interband velocity is schematically depicted in Fig.~\ref{fig:transitions}. 

To be concrete, we now focus on the nonunitary helical $p$-wave state described by Eq.~\eqref{eq:mixedHelical}. Since the pairing takes place between equal pesudospins, the two pseudospin sectors are decoupled and can be discussed separately. When inter-band pairing is neglected, in the Nambu spinor basis $(c_{i\bk s},c^\dagger_{i\bar{\bk}s})^T$, we can define $H^\text{BdG}_{\bk s} |i\bk s\rangle = E_{i\bk s}|i\bk s\rangle$ with $E_{i\bk s} = \sqrt{(\epsilon_{i\bk})^2 + |\Delta^{ss}_{i\bk}|^2} $ and $|i\bk s\rangle = (u_{i\bk s}, v_{i\bk s})^T$, and let $|\bar{i}\bk s\rangle = (-v^\ast_{i\bk s}, u^\ast_{i\bk s})^T$ be the particle-hole conjugate contourpart. Consider the quantity $Q^{ab}_{xy,\bk}$ originating from the following virtual processes,
\begin{eqnarray}
&& |\bar{1}\bk s\rangle \xrightarrow{ \mathcal{V}^{12}_{y\bk s}} |2\bk s\rangle  \xrightarrow{ \mathcal{V}^{21}_{x\bk s}} |\bar{1}\bk s\rangle ~~~ - ~~~(x \leftrightarrow y)\,,  \nonumber
\end{eqnarray}
for which a straightforward derivation leads to
\begin{eqnarray}
&&u_{2\bk s}^\ast v_{1\bk s}^\ast u_{1\bk s} v_{2\bk s} [ (V^{12}_{x \bar{\bk} s})^\ast V^{21}_{y \bk s} -(V^{12}_{y \bar{\bk} s})^\ast V^{21}_{x\bk s}] \nonumber \\
+ && v_{2\bk s}^\ast u_{1\bk s}^\ast v_{1\bk s} u_{2\bk s} [V^{12}_{x\bk s} (V^{21}_{y\bar{\bk}s})^\ast - V^{12}_{y\bk s} (V^{21}_{x\bar{\bk} s})^\ast  ] \nonumber \\
+ &&|u_{1\bk s}|^2|v_{2\bk s}|^2[(V^{12}_{x \bar{\bk} s})^\ast (V^{21}_{y \bar{\bk} s})^\ast-(V^{12}_{y \bar{\bk} s})^\ast (V^{21}_{x \bar{\bk} s})^\ast] \nonumber \\
+ &&  |u_{2\bk s}|^2|v_{1\bk s}|^2(V^{12}_{x\bk s}V^{21}_{y\bk s}-V^{12}_{y\bk s} V^{21}_{x\bk s})  \,.
\end{eqnarray}
This expression can be significantly simplified in the weak-coupling limit. In a multiband system whose bands are not degenerate at generic wavevectors, the weak pairing gap is typically much smaller than the normal-state band separation. Hence, it is reasonable to take the pairing to be separately finite on one of the bands and negligible on the other. In this approximation, only the last two lines of the above expression contribute. Let's assume, for example, that only band-1 superconducts, then we obtain~\cite{supplementary}
\begin{eqnarray}
 - i\rho_{1\bk s} (\epsilon_{1\bk}-\epsilon_{2\bk})^2  \mathcal{B}^1_{xy,\bk s}  \,,  
\label{eq:rhoBerry}
\end{eqnarray}
where $\rho_{i\bk s} = |v_{i\bk s}|^2$ is the spin occupancy at zero temperature. It is worth stressing that the spin Berry curvature $\mathcal{B}^1_{xy,\bk s}$ is a normal-state quantity associated with the Bloch electrons, which is completely different from the Berry curvature of Bogoliubov quasiparticles; the latter is a notion widely used in the context of topological superconducting states. 

\begin{figure}
\includegraphics[width=8.5cm]{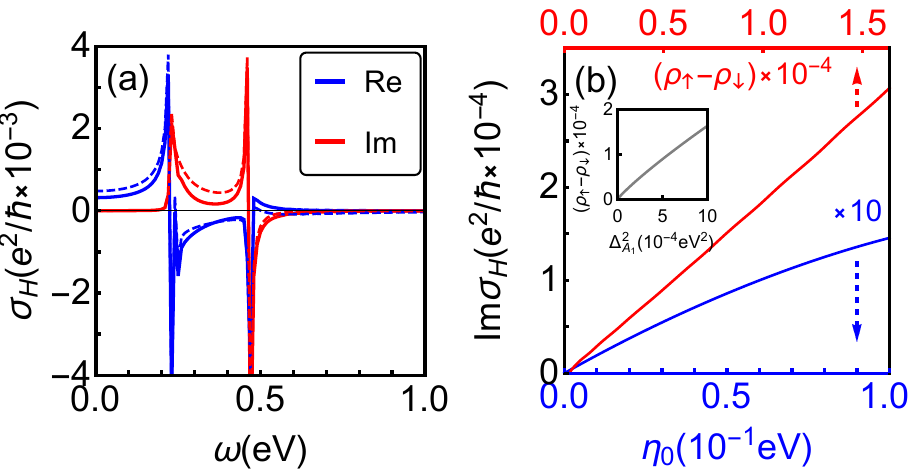}
\caption{(a) The zero-temperature Hall conductivity as a function of $\omega$ for the $A_{1u}+iA_{2u}$ state. The blue and red curves represent the real and imaginary part of the Hall conductivity, respectively. The solid curves are evaluated with intraband-only pairings while the dashed curves are obtained with both intra- and inter-band pairings present. We take $\Delta_{A_1}=2\Delta_{A_2}=0.01$ eV. (b) The imaginary part of the Hall conductivity as a function of $\eta_0$ (blue) and of the spin polarization $\rho_{\uparrow}-\rho_{\downarrow}$ (red), where $\rho_{s}=\sum_{i,\bk} \rho_{i\bk s}$. The conductivity is evaluated at a fixed frequency $\omega_0=0.3$ eV. The blue curve is calculated with $\Delta_{A_1}=2\Delta_{A_2}=0.01$ eV, and the result has been scaled up by a factor of 10 for better visualization. The red curve is obtained with $\eta_0=0.1$ eV and with varying overall pairing gap magnitude, while keeping the ratio $\Delta_{A_1}/\Delta_{A_2}=2$ fixed. The inset shows the spin polarization as a function of $\Delta_{A_1}^2$. Other parameters not mentioned here are the same as in Fig.~\ref{fig:Berry} (a).}
\label{fig:cond}
\end{figure}

Since the two pseudospin states feature opposite Berry curvature and since their occupancy $\rho_{i\bk s}$ differs because $|\Delta^{\ua\ua}_{i\bk}| \neq |\Delta^{\da\da}_{i\bk}|$, from Eq.~\eqref{eq:rhoBerry} it is now clear how a charge Hall response may arise in our model. The frequency-dependent Hall conductivity $\sigma_H$ for our two-band model with a representative set of parameters is plotted in Fig.~\ref{fig:cond} (a). As explained in detail in the Supplementary~\cite{supplementary}, the pairing is first constructed in the original orbital basis and then transformed into the band basis in which the kinetic part of the BdG Hamiltonian is diagonal. A model with only intraband pairing is then obtained by purposely removing the interband part. Numerical results both with and without interband pairing are presented in Fig.~\ref{fig:cond} (a).  As one can see, with finite SOC, whether interband pairing is included or not does not qualitatively change the conclusion. On the other hand, in the absence of SOC, $\sigma_H$ vanishes completely if the interband pairing is absent (not shown). 

A notable feature in the imaginary part of $\sigma_H$ is the low-frequency cutoff that is of the order of normal-state interband energy separation. This further confirms that only interband scattering processes contribute to the intrinsic Hall response and attests to the critical role of interband velocity. In the Supplementary~\cite{supplementary}, we perform another model calculation, wherein the pairing is constructed directly in the band basis, and obtain qualitatively similar results. 

Figure~\ref{fig:cond} (b) shows the variation of $\sigma_{H}$ with the SOC strength $\eta_0$ and also with the superconducting-state spin polarization $\rho_\ua - \rho_\da$. Consistent with the above analyses, $\sigma_{H}$ increases roughly linearly with $\eta_0$ (thus the normal-state spin Hall conductivity, see Fig.~\ref{fig:Berry} (b)) and the spin polarization. In practice, the spin polarization is tuned by varying the overall gap amplitude while keeping their ratio fixed with $\Delta_{A_1}/\Delta_{A_2}=2$ [see inset of Fig.~\ref{fig:cond} (b)]. 

For the 3D chiral $p$-wave state, the Hamiltonian in general cannot be block-diagonalized into individual pseudospin subspace. One exception is when only the out-of-plane pairing $\Delta_\perp$ is present in the gap function Eq.~\eqref{eq:3DchiralPwave}. Finite charge Hall effect is hence also expected, following the above analyses. An exemplary calculation for a representative 3D chiral $p$-wave state is provided in the Supplementary~\cite{supplementary}. The results are similar to those obtained for the nonunitary helical $p$-wave model. Further taking into account in-plane pairing $\Delta_{\parallel}$ does not change this conclusion qualitatively. Thus, $\sigma_H$ is predominantly determined by $\Delta_\perp$, which sets the degree of the non-unitarity of the pairing. 

{\bf Summary and final remarks --} We have shown that superconducting \SRO~without interband Cooper pairing can support intrinsic anomalous charge Hall response and polar Kerr effect, if it condenses into one of the nonunitary odd-parity pairing states. In such case, the Kerr rotation angle at the experimental photon energy of $\hbar\omega=0.8$ eV shall be larger for samples with higher quality and shall remain finite even for pristine samples. The charge Hall conductivity is determined by both the nonunitary-pairing-induced spin polarization and the SOC-derived spin Berry curvature of the normal-state Bloch electrons. This mechanism applies even when only one of the bands crosses the Fermi energy and develops Cooper pairing, and may also be applied to other superconductors where Kerr effect has been reported. 

To make connection with the Kerr measurement in \SRO, we evaluate the charge Hall conductivity of our two-band model with band parameters and pairing gaps consistent with realistic estimates~\cite{Sharma:20} and estimate the Kerr rotation angle. Details are given in the Supplementary~\cite{supplementary}. For the 3D chiral $p$-wave state with $\Delta_{\parallel}=0.35$ meV and for $\Delta_{\perp}$ ranging from 0 to 0.35 meV, one obtains Kerr rotation $\theta_K \in (0, 2.8)$ nrad. Given the neglect of the third band and the crudeness of our approximation, it is unclear whether this state can explain the experimental value of $\theta_{K} \approx 60$ nrad~\cite{Xia:06} observed at low temperatures. On the other hand, for the $A_{1u}+iA_{2u}$ state, taking $\Delta_{A_{1}} = 0.35$ meV and for $\Delta_{A_{2}} \in (0, 0.35)$ meV, we obtain $\theta_K \in (0, 25.6)$ nrad, which is closer to the experimental result in the limit $\Delta_{A_{1}} = \Delta_{A_{2}}$ -- as would be the case if the two order parameters are nearly degenerate.  

{\bf Acknowledgements --} We acknowledge helpful discussions with Wen Sun. This work is supported by NSFC under grant No.~11904155 and No.~12374042, the Guangdong Science and Technology Department under Grant 2022A1515011948, a Shenzhen Science and Technology Program (Grant No.~KQTD20200820113010023), the Guangdong Provincial Key Laboratory under Grant No.~2019B121203002, and the Fundamental Research Funds for the Central Universities (Grant No. E2E44305). Z. W. is supported by the James Franck Institute at the University of
Chicago. Computing resources are provided by the Center for Computational Science and Engineering at Southern University of Science and Technology.

\clearpage
\appendix
\setcounter{equation}{0}
\setcounter{figure}{0}
\renewcommand {\theequation} {S\arabic{equation}}
\renewcommand {\thefigure} {S\arabic{figure}}
\begin{widetext}

\section{Supplementary Material for: Quantum-geometry-induced anomalous Hall effect in nonunitary superconductors and application to \SRO}

\section{I. General formalism}
In this section, we introduce the formalism for evaluating the anomalous (charge) Hall conductivity (AHC) in the absence interband pairing. Typically, one first writes down the model Hamiltonian constructed in orbital basis and then transform it into band basis wherein the kinetic part of the Hamiltonian is diagonalized. In accordance, the velocity operator and the pairing matrix should also be transformed into the band basis description. Below, we shall exemplify our method with a two-orbital model with $d_{xz}$ and $d_{yz}$ orbitals residing on each site of a square lattice.

The Nambu spinor in orbital basis is written as
\begin{align}
\Psi_{\bk}^{\dagger} &= (\psi^{\dagger}_{\bk}, \psi^{T}_{\bar{\bk}}) \,,  \\
\psi^{\dagger}_{\bk} = (c^{\dagger}_{x\bk\uparrow},c^{\dagger}_{x\bk\downarrow},c^{\dagger}_{y\bk\uparrow}, c^{\dagger}_{y\bk\downarrow})  \,,  & \qquad
\psi^{T}_{\bar{\bk}} = (c_{x\bar{\bk}\uparrow},c_{x\bar{\bk}\downarrow},c_{y\bar{\bk}\uparrow},c_{y\bar{\bk}\downarrow})  \,.  \notag
\end{align}
Where $\bar{\bk}\equiv-\bk$ and subscript $x,y$ represents $d_{xz}$, $d_{yz}$ orbital, respectively. The normal state Hamiltonian $H_{0\bk}$ is given in the maintext and the corresponding Bogoliubov-de Gennes (BdG) Hamiltonian matrix in the orbital basis is
\begin{equation}
H^\text{orb}_{\bk}=\begin{pmatrix}
H_{0\bk} & \Delta^\text{orb}_{\bk}  \\
(\Delta^\text{orb}_{\bk})^{\dagger}  &  -H^{T}_{0\bar{\bk}}
\end{pmatrix}   \,.
\end{equation}
Let $U_{\bk}$ be a unitary matrix that diagonalizes $H_{0\bk}$, {\it i.e.}, $U^{-1}_{\bk} H_{0\bk} U_{\bk}=  \Gamma_{\bk}$, where $\Gamma_{\bk}=\text{diag} (
\epsilon_{1\bk}, \epsilon_{1\bk}, \epsilon_{2\bk}, \epsilon_{2\bk})$. Then we have the band-basis BdG Hamiltonian given by 
\begin{equation}
H^\text{BdG}_{\bk} \equiv
\label{Eq:transformation}
\begin{pmatrix}
U^{-1}_{\bk}  &  \\
 & \bar{U}^{-1}_{\bk}
\end{pmatrix}
H^\text{orb}_{\bk}
\begin{pmatrix}
U_{\bk}  &  \\
 & \bar{U}_{\bk}
\end{pmatrix}
=
\begin{pmatrix}
\Gamma_{\bk} & \Delta_{\bk}  \\
\Delta_{\bk}^{\dagger} & -\Gamma_{\bk}
\end{pmatrix}  \,.
\end{equation}
Here $\bar{U}_{\bk}=U^{*}_{\bar{\bk}}$. Here, the pairing matrix in band basis is
\begin{equation}
\label{Eq:deltrans}
\Delta_{\bk}= U^{-1}_{\bk} \Delta^\text{orb}_{\bk} \bar{U}_{\bk} \,.
\end{equation}
Multi-orbital models with the pairing constructed in the orbital basis generically exhibit finite interband pairing after the transformation procedure described above. A model with only intraband pairing can be obtained by removing the interband pairing by hand. We denoted the pairing matrix with only intraband pairings by $\tilde{\Delta}_{\bk}$, and the corresponding reduced BdG Hamiltonian is
\begin{equation}
\tilde{H}^\text{BdG}_{\bk} =
\begin{pmatrix}
\Gamma_{\bk} & \tilde{\Delta}_{\bk}  \\
\tilde{\Delta}_{\bk}^{\dagger} & -\Gamma_{\bk}
\end{pmatrix}  \,,
\end{equation}
and its Green's function in spectral representation reads
\begin{equation}
\tilde{G}(i\omega_n,\bk)\equiv \left( i\omega_n-\tilde{H}^\text{BdG}_{\bk}\right)^{-1}=\sum_{a\bk} \frac{|a\bk\rangle \langle a\bk|}{i\omega_n-E_{a\bk}} \,.
\end{equation} 
The velocity operator in orbital basis is defined as 
\begin{equation}
\mathcal{V}^\text{orb}_{\mu\bk}= \begin{pmatrix}
\partial_{\mu} H_{0\bk} & \\
 &  \partial_{\mu} (-H_{0\bar{\bk}}^{T} )
\end{pmatrix}
=\begin{pmatrix}
V^\text{orb}_{\mu\bk} & \\
 &  -(V^\text{orb} _{\mu\bar{\bk}})^T
\end{pmatrix}  \,.
\end{equation}
The transformation of the velocity operator follows the same way of transformation of $H_{\bk}^{BdG}$. In band basis it reads
\begin{equation}
\label{Eq:vtrans}
\mathcal{V}_{\mu\bk}= \begin{pmatrix}
V_{\mu\bk}  & \\
 & -(V_{\mu\bar{\bk}})^{T}
\end{pmatrix} 
=\begin{pmatrix}
U^{-1}_{\bk}  &  \\
 & \bar{U}^{-1}_{\bk}
\end{pmatrix}
\mathcal{V}^\text{orb}_{\mu\bk}
\begin{pmatrix}
U_{\bk}  &  \\
 & \bar{U}_{\bk}
\end{pmatrix}
= \begin{pmatrix}
U^{-1}_{\bk} V^\text{orb}_{\mu\bk} U_{\bk} &  \\
 & -\bar{U}^{-1}_{\bk} (V^\text{orb}_{\mu\bar{\bk}})^T \bar{U}_{\bk}
\end{pmatrix}  \,.
\end{equation}
Note that both $V^\text{orb}_{\mu\bk}$ and $V_{\mu\bk}$ are Hermitian matrices. The AHC at one-loop approximation expressed in band basis is given by
\begin{align}
\label{Eq:sigmaH_G}
\sigma_{H}(\omega)/(\frac{e^2}{\hbar}) &= \frac{i T}{4N\omega}\sum_{\bk,\omega_n} \text{Tr}\left[ \mathcal{V}_{x\bk} \tilde{G}(i\omega_n,\bk) \mathcal{V}_{y\bk}\tilde{G}(i\omega_n+\omega,\bk)  - (x\leftrightarrow y)  \right]  \\
\label{Eq:sigmaH_spec}
&= \frac{i}{4N\omega}\sum_{\bk,ab} \frac{f(E_{a\bk})-f(E_{b\bk})}{\hbar\omega+i0^{+} +E_{a\bk}-E_{b\bk}} \left[ \langle b\bk|\mathcal{V}_{x\bk} |a\bk \rangle \langle a\bk | \mathcal{V}_{y\bk}|b\bk \rangle  - (x\leftrightarrow y)  \right] \,.
\end{align}

\section{II. Mixed helical $p$-wave states}
In this section, we present the modeling of two mixed helical $p$-wave states, {\it i.e.}, $A_{1u}+iA_{2u}$ and $B_{1u}+iB_{2u}$, and the numerically obtained AHC. In both of these states, Cooper pairing develops between equal (pseudo-)spin electrons, although the pairing amplitude differs for the two spin species. Our following analyses will be based on the $d_{xz}$-$d_{yz}$ two-orbital whose Hamiltonian is given by Eq.~(5) in the maintext, $H_{0\bk}=\epsilon_{\bk}+\tilde{t}_{\bk}\sigma_z+\lambda_{\bk}\sigma_x+\eta_{\bk}\sigma_y  s_z$. In this case the spin-up and spin-down species can be treated separately, hence the calculation of AHC is analytically tractable. 

Here we present the zero-temperature derivation of the AHC. At zero temperature, the only transition processes that contribute to the AHC are the ones that start from a negative energy state, and then scatter into a positive state via interband velocity and finally return to the original state. In this case the AHC can be obtained by solving the BdG Hamiltonian directly and substituting the eigenfunctions into the spectral representation. For the spin-up sector, the normal-state Hamiltonian is
\begin{equation} 
H_{0\bk\uparrow} = \begin{pmatrix}
\epsilon_{x\bk} &  \lambda_{\bk} - i\eta_{\bk}   \\
\lambda_{\bk} + i\eta_{\bk}  &  \epsilon_{y\bk} 
\end{pmatrix}  \,.
\end{equation} 
where 
\begin{align}
\epsilon_{x\bk} &= -t_{+}\cos k_{x}-t_{-} \cos k_{y} -\mu \,,  &\lambda_{\bk} &=\lambda \sin k_x \sin k_y \,,  \notag \\
\epsilon_{y\bk} &= -t_{-} \cos k_{x}-t_{+} \cos k_{y} -\mu  \,,  &\eta_{\bk} &= \eta_0+\eta_1(\cos k_x+\cos k_y)  \,,
\end{align}
where $t_{\pm}=t \pm \tilde{t}$. $t$ and $\tilde{t}$ are hopping parameters in $\epsilon_{\bk}$ and $\tilde{t}_{\bk}$, respectively. The unitary matrix that diagonalize $H_{0\bk\uparrow}$ is
\begin{equation}
\label{Eq:umat}
U_{\bk\uparrow} = \begin{pmatrix}
-e^{-i\phi_{\bk}}q_{\bk} &   p_{\bk}  \\
 p_{\bk}  &  q_{\bk} e^{i\phi_{\bk}} 
\end{pmatrix}  \,,
\end{equation}
where
\begin{align*}
p_{\bk} &= \sqrt{\frac{1}{2}\left(1+\frac{\epsilon_{x\bk}-\epsilon_{y\bk}}{\xi_{\bk}}\right)}  \,,\quad
& q_{\bk} &= \sqrt{\frac{1}{2}\left(1-\frac{\epsilon_{x\bk}-\epsilon_{y\bk}}{\xi_{\bk}}\right)}  \,, \\
\xi_{\bk} &= \sqrt{(\epsilon_{x\bk}-\epsilon_{y\bk})^2+4(\lambda_{\bk}^2+\eta_{\bk}^2)} \,,\quad
&e^{i\phi_{\bk}} &=\frac{\lambda_{\bk}+ i \eta_{\bk}}{\sqrt{\lambda_{\bk}^2+\eta_{\bk}^2}} \,.
\end{align*}
And
\begin{equation}
U^{-1}_{\bk\uparrow} H^{\uparrow}_{0\bk} U_{\bk\uparrow} = \Gamma_{\bk\uparrow}=\begin{pmatrix}
\epsilon_{1\bk} &  \\
& \epsilon_{2\bk}  
\end{pmatrix} \,,
\end{equation}
where the Bloch spectrum are given by $\epsilon_{1/2\bk} = \frac{1}{2}(\epsilon_{x\bk}+\epsilon_{y\bk} \mp \xi_{\bk})$. With the unitary matrix obtained, we can transform the velocity operator and pairing matrix into band basis. The velocity operator in orbital basis is
\begin{align}
V^\text{orb}_{\mu\bk\uparrow} = \partial_{\mu} H_{0\bk\uparrow} 
\equiv  \begin{pmatrix}
V^{\text{orb},11}_{\mu\bk\uparrow} & V^{\text{orb},12}_{\mu\bk\uparrow}   \\
V^{\text{orb},21}_{\mu\bk\uparrow} & V^{\text{orb},22}_{\mu\bk\uparrow}
\end{pmatrix} \,.
\end{align}
Transform into band basis, (cf. Eq.~\eqref{Eq:vtrans} for the transformation of full velocity operator), we have
\begin{equation}
V_{\bk\uparrow} = U_{\bk\uparrow}^{-1} V^\text{orb}_{\mu\bk\uparrow} U_{\bk\uparrow} =   \begin{pmatrix}
V_{\mu\bk\uparrow}^{11}  & V_{\mu\bk\uparrow}^{12}  \\
V_{\mu\bk\uparrow}^{21}   & V_{\mu\bk\uparrow}^{22} 
\end{pmatrix} \,.
\end{equation}
By using Eq.~\eqref{Eq:umat}, the explicit expression for interband velocity is
\begin{align}
V_{\mu\bk\uparrow}^{12} &= -e^{2i\phi_{\bk}}q_{\bk}^2 V^{\text{orb},12}_{\mu\bk\uparrow}+p_{\bk}^2 V^{\text{orb},21}_{\mu\bk\uparrow}+ e^{i\phi_{\mu\bk}}p_{\bk}q_{\bk} (V^{\text{orb},22}_{\mu\bk\uparrow}-V^{\text{orb},11}_{\mu\bk\uparrow})  \,.
\end{align}
The pairing matrix in band basis could be obtained via Eq.~\eqref{Eq:deltrans},
\begin{equation}
\Delta_{\bk\uparrow} = U_{\bk\uparrow}^{-1}\Delta^\text{orb}_{\bk} U_{\bar{\bk}\uparrow}^{*}  \rightarrow  \tilde{\Delta}_{\bk\uparrow}=
\begin{pmatrix}
\Delta_{1\bk\uparrow}   & \\
& \Delta_{2\bk\uparrow} 
\end{pmatrix}  \,,
\end{equation}
where we have eliminated the interband pairings in the last step. We shall discuss the explicit form of $\Delta_{i\bk\uparrow}$, $i=1,2$ in the end of this section. The reduced BdG Hamiltonian then reads
\begin{equation}
\label{Eq:hbdgup}
\tilde{H}^\text{BdG}_{\bk\uparrow}  = \begin{pmatrix}
\Gamma_{\bk\uparrow}    & \Delta_{\bk\uparrow}  \\
\Delta_{\bk\uparrow}^{\dagger}  &  -\Gamma_{\bk\uparrow}  \\
\end{pmatrix}  \,.
\end{equation}
Diagnalizing the above Hamiltonian and plugging the eigenstates into the spectral representation of the AHC, see Eq.~\eqref{Eq:sigmaH_spec}, at zero temperature we obtain
\begin{align}
\label{Eq:ahcup}
\sigma^{\uparrow}_{H} (\omega)/(\frac{e^2}{\hbar}) &= \frac{i}{2N}\sum_{\bk}\left[  \frac{ \epsilon_{1\bk}/E_{1\bk\uparrow}-\epsilon_{2\bk}/E_{2\bk\uparrow}}{(\hbar\omega+i0^{+})^2-(E_{1\bk\uparrow}+E_{2\bk\uparrow})^2} \right]  \left[ V_{x\bk\uparrow}^{12} V_{y\bk\uparrow}^{21} - V_{y\bk\uparrow}^{12} V_{x\bk\uparrow}^{21} \right]  \,,  \\
&= \frac{1}{2N}\sum_{\bk}\left[  \frac{ \epsilon_{1\bk}/E_{1\bk\uparrow}-\epsilon_{2\bk}/E_{2\bk\uparrow}}{(\hbar\omega+i0^{+})^2-(E_{1\bk\uparrow}+E_{2\bk\uparrow})^2} \right] (\epsilon_{1\bk\uparrow}-\epsilon_{2\bk\uparrow})^2 \mathcal{B}^{1}_{xy,\bk\uparrow}
\end{align}
where the BdG energy spectrum is given by $ E_{i\bk \uparrow}=\sqrt{(\epsilon_{i\bk })^2+|\Delta_{i\bk \uparrow}|^2}$, $i=1,2$. From Eq.~\eqref{Eq:ahcup} we can explicitly see that the AHC is directly related the interband velocity and thus the Berry curvature of the normal state. We emphasize that this relation is a common feature in this mechanism of generating the intrinsic AHC.

The AHC for spin-down sector can be obtained by following the exactly same procedure and the result is
\begin{equation}
\label{Eq:ahcdown}
\sigma^{\downarrow}_{H} (\omega)/(\frac{e^2}{\hbar}) =\frac{1}{2N}\sum_{\bk}\left[  \frac{ \epsilon_{1\bk}/E_{1\bk\downarrow}-\epsilon_{2\bk}/E_{2\bk\downarrow}}{(\hbar\omega+i0^{+})^2-(E_{1\bk\downarrow}+E_{2\bk\downarrow})^2} \right](\epsilon_{1\bk\downarrow}-\epsilon_{2\bk\downarrow})^2 \mathcal{B}^{1}_{xy,\bk\downarrow}  \,.
\end{equation}
The total charge AHC, which is the sum of AHC for two spin species thus given by
\begin{equation}
\label{Eq:cond}
\sigma_{H} (\omega) /(\frac{e^2}{\hbar})= \frac{1}{2N}\sum_{\bk}\left[  \frac{ \epsilon_{1\bk}/E_{1\bk\uparrow}-\epsilon_{2\bk}/E_{2\bk\uparrow}}{(\hbar\omega+i0^{+})^2-(E_{1\bk\uparrow}+E_{2\bk\uparrow})^2}-   \frac{ \epsilon_{1\bk}/E_{1\bk\downarrow}-\epsilon_{2\bk}/E_{2\bk\downarrow}}{(\hbar\omega+i0^{+})^2-(E_{1\bk\downarrow}+E_{2\bk\downarrow})^2}  \right] (\epsilon_{1\bk}-\epsilon_{2\bk})^2 \mathcal{B}^{1}_{xy,\bk}   \,,
\end{equation}
where we have used the time-reversal symmetry of the normal state,
\begin{align}  
\epsilon_{i\bk \uparrow} &= \epsilon_{i\bar{\bk} \downarrow} \equiv \epsilon_{i\bk}  \quad i=1,2  \,,  \\
\mathcal{B}^{1}_{xy,\bk\uparrow} &=- \mathcal{B}^{1}_{xy,\bar{\bk}\downarrow}  \equiv
\mathcal{B}^{1}_{xy,\bk}  \,.
\end{align}

We now give the explicit form of pairing matrix. In orbital basis, the pairing matrix of $A_{1u}+iA_{2u}$ state is given by ~\cite{Huang2021prr},
\begin{align}
\Delta_{\bk} &=   \Delta_{A_{1}}\left[ \sin k_{x} \frac{\sigma_0+\sigma_z}{2} s_{x}+\sin k_{y}\frac{\sigma_0-\sigma_z}{2}  s_{y} \right]is_y+ i \Delta_{A_{2}}\left[ \sin k_{x}\frac{ \sigma_0+\sigma_z}{2} s_{y}-\sin k_{y}\frac{\sigma_0-\sigma_z}{2}  s_{x}\right] is_{y} \notag   \\
&=
\begin{pmatrix}
-(\Delta_{A_{1}}+\Delta_{A_{2}})\sin k_x  & & & \\
& (\Delta_{A_{1}}-\Delta_{A_{2}})\sin k_x & &    \\
& & i(\Delta_{A_{1}}+\Delta_{A_{2}})\sin k_y & \\
& & & i(\Delta_{A_{1}}-\Delta_{A_{2}})\sin k_y
\end{pmatrix}  \,,
\end{align}
where $\sigma_i (s_i) \,,  i=1,2,3$ are Pauli matrices operating in orbital (spin) manifold. Here we consider the basis functions with only intraorbital pairing. Turning into band basis, the corresponding intraband pairings are given by
\begin{align}
\Delta_{1\bk\uparrow}  &= -(\Delta_{A_1}+\Delta_{A_2})(e^{2i\phi_{\bk}} q_{\bk}^2 \sin k_x -i p_{\bk}^2 \sin k_y )  \,, \notag  \\
\Delta_{2\bk\uparrow} &=  -(\Delta_{A_1}+\Delta_{A_2})(p_{\bk}^2 \sin k_x - ie^{-2i\phi_{\bk}} q_{\bk}^2 \sin k_y) \,,  \\
\Delta_{1\bk\downarrow} &= (\Delta_{A_1}-\Delta_{A_2})(e^{-2i\phi_{\bk}} q_{\bk}^2 \sin k_x +i p_{\bk}^2 \sin k_y )   \,,  \notag  \\
\Delta_{2\bk\downarrow} &=  (\Delta_{A_1}-\Delta_{A_2})(p_{\bk}^2 \sin k_x + ie^{2i\phi_{\bk}} q_{\bk}^2 \sin k_y) \,.  \notag  
\end{align}
Around $\Gamma$ point, we have $p_{\bk}^2=q_{\bk}^2\approx \frac{1}{2}$, $e^{i\phi_{\bk}}\approx i$, the intraband pairings take the form of simplest chiral $p$-wave with chirality $\pm 1$,
\begin{align}
\label{Eq:delb}
\Delta_{1\bk\uparrow}  &\approx  -(\Delta_{A_1}+\Delta_{A_2})(k_{x} +i k_{y})  \,,  \notag  \\
\Delta_{2\bk\uparrow}  &\approx  -(\Delta_{A_1}+\Delta_{A_2})(k_{x} +i  k_{y}  ) \,, \\
\Delta_{1\bk\downarrow} &\approx (\Delta_{A_1}-\Delta_{A_2}) (k_{x} -i  k_{y} ) \,,  \notag  \\
\Delta_{2\bk\downarrow} &\approx (\Delta_{A_1}-\Delta_{A_2}) (k_{x} -i  k_{y})  \,. \notag  
\end{align}
Note that in terms of total angular momentum eigenvalues $J=L+S$, the above four Bloch bands can be labelled by $j_{z}=\frac{3}{2}, -\frac{1}{2}, -\frac{3}{2}$ and $ \frac{1}{2}$, respectively. The symmetries of intraband pairings are in line with the symmetry analysis in Ref.~\cite{Chen:21}.

We can also construct the model directly in band basis using the intraband pairings given by Eq.~\eqref{Eq:delb} (note that in lattice model $k_{x(y)}$ is replaced by $\sin k_{x(y)}$). The numerical result of $\omega$-dependent Hall conductivity $\sigma_{H}$ using analytical expression Eq.~\eqref{Eq:cond} is present in Fig.~\ref{fig:condhelical}. The result is qualitatively in agreement with that calculated from orbital basis, see Fig.3(a) in maintext.

\begin{figure}
\includegraphics[width=8cm]{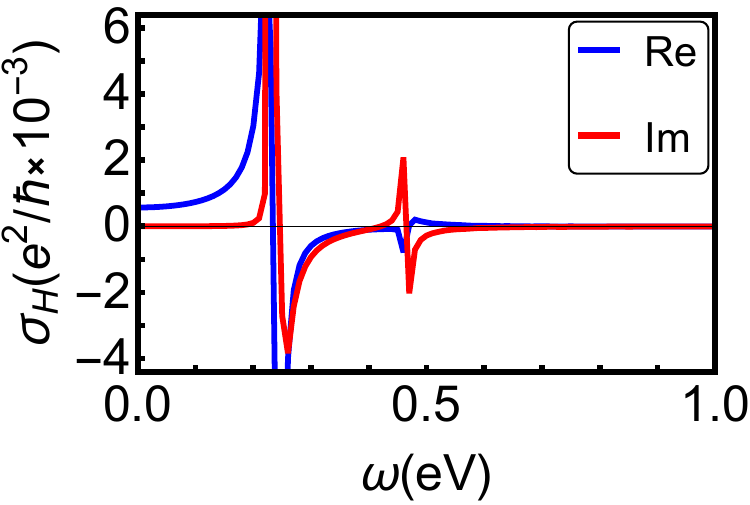}
\caption{The Hall conductivity as a function of $\omega$ for $A_{1u}+A_{2u}$ state.  The calculation is performed in the band basis with only intraband pairings, see Eq.~\eqref{Eq:delb}. The gap amplitudes for spin-up and -down bands are given by $\Delta_{\uparrow}=\Delta_{A_{1u}}+\Delta_{A_{2u}}=0.01$ eV $\Delta_{\downarrow}=\Delta_{A_{1u}}-\Delta_{A_{2u}}=0.005$ eV.}
\label{fig:condhelical}
\end{figure}

In the case of $B_{1u}+iB_{2u}$ state, the pairing matrix in orbital basis is
\begin{align}
\Delta_{\bk} &=  \Delta_{B_{1}}\left[ \sin k_{x} \frac{\sigma_0+\sigma_z}{2} s_{x}-\sin k_{y}\frac{\sigma_0-\sigma_z}{2}  s_{y} \right]is_y+ i \Delta_{B_{2}}\left[ \sin k_{x}\frac{ \sigma_0+\sigma_z}{2} s_{y}+\sin k_{y}\frac{\sigma_0-\sigma_z}{2}  s_{x}\right] is_{y} \notag   \\
&=
\begin{pmatrix}
 -(\Delta_{B_{1}}+\Delta_{B_{2}})\sin k_{x}  &   & & \\
   & (\Delta_{B_{1}}-\Delta_{B_{2}})\sin k_{x}   & & \\
  & & -i(\Delta_{B_{1}}+\Delta_{B_{2}})\sin k_{y}   &   \\
  & &  &  -i(\Delta_{B_{1}}-\Delta_{B_{2}})\sin k_{y} 
\end{pmatrix}  \,,
\end{align}
and the corresponding intraband pairings are
\begin{align}
\Delta_{1\bk\uparrow}  &= -(\Delta_{B_1}+\Delta_{B_2})(e^{2i\phi_{\bk}} q_{\bk}^2 \sin k_x + i p_{\bk}^2 \sin k_y )  \,, \notag  \\
\Delta_{2\bk\uparrow} &=  -(\Delta_{B_1}+\Delta_{B_2})(p_{\bk}^2 \sin k_x + ie^{-2i\phi_{\bk}} q_{\bk}^2 \sin k_y)  \,,  \\
\Delta_{1\bk\downarrow} &= (\Delta_{B_1}-\Delta_{B_2})(e^{-2i\phi_{\bk}} q_{\bk}^2 \sin k_x -i p_{\bk}^2 \sin k_y ) \,, \notag  \\
\Delta_{2\bk\downarrow} &=  (\Delta_{B_1}-\Delta_{B_2})(p_{\bk}^2 \sin k_x - ie^{2i\phi_{\bk}} q_{\bk}^2 \sin k_y) \,.  \notag  
\end{align}
We could see that the form of $B_{1u}+iB_{2u}$ pairing is essentially equivalent to that of $A_{1u}+iA_{2u}$ state but with opposite chirality.

\section{III. Three-dimensional (3D) chiral $p$-wave}
In a spin-orbit-coupled model, the $E_{u}$ chiral $p$-wave pairing channel shall generically acquire both in-plane and out-of-plane pairing components: $\Delta_{||}(k_{x}+ik_{y})\hat{z}$ and $\Delta_\perp(\hat{x}+i\hat{y})k_{z}$. In the current two-orbital model, the simplest orbital-basis 3D chiral $p$-wave pairing reads,
\begin{align}
\Delta_{\bk} &=  \Delta_{\parallel}\left[ \frac{\sigma_0+\sigma_z}{2}\sin k_{x} s_{z}+i\frac{\sigma_0-\sigma_z}{2} \sin k_{y} s_{z} \right]is_{y} + \Delta_{\perp}\sin k_z\left[ \frac{\sigma_0+\sigma_z}{2} s_{x}+i\frac{\sigma_0-\sigma_z}{2} s_{y} \right]is_{y} \notag   \\
&=
\label{Eq:del3d}
\begin{pmatrix}
 -\Delta_{\perp}\sin k_{z}  &  \Delta_{\parallel}\sin k_{x}  & & \\
 \Delta_{\parallel}\sin k_{k}  &  \Delta_{\perp}\sin k_{z}  & & \\
  & & -\Delta_{\perp}\sin k_{z}  &  i\Delta_{\parallel}\sin k_{y}  \\
  & & i\Delta_{\parallel}\sin k_{y}  &  -\Delta_{\perp}\sin k_{z}   
\end{pmatrix}  \,.
\end{align}
In this model the spin-up and -down degrees of freedom are no longer decoupled. Thus we can only calculated the AHC numerically using the spectral representation in Eq.~\eqref{Eq:sigmaH_spec}. The $\omega$-dependent Hall conductivity is shown in Fig.~\ref{fig:cond3d} which is similar to those obtained in helical $p$-wave models.

\begin{figure}
\includegraphics[width=8cm]{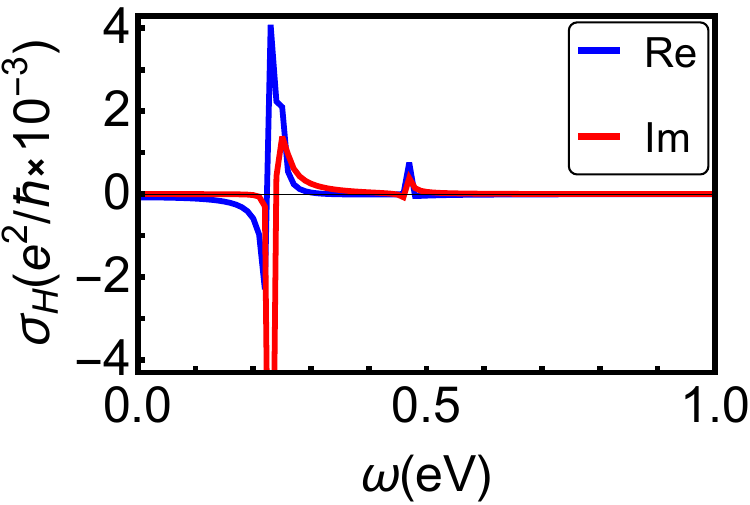}
\caption{The Hall conductivity as a function of $\omega$ for 3D chiral $p$-wave state. The calculation is based on the model in orbital basis, see Eq.~\eqref{Eq:del3d}. The interband  pairings are eliminated after transforming into band basis. The gap amplitudes are $\Delta_{\parallel}=\Delta_{\perp}=0.01$ eV.}
\label{fig:cond3d}
\end{figure}

To further investigate the relationship between the AHC and spin polarization, we calculated the $\Delta_{\perp(\parallel)}^{2}$-dependence of $\sigma_{H}$, see Fig.~\ref{fig:conddel}. Note that in this model the spin polarization is controlled by the gap amplitude of the out-of-plane pairing $\rho_{\uparrow}-\rho_{\downarrow}\propto \Delta_{\perp}^{2}$ (not shown here). Thus we expect that $\sigma_{H}$ is predominated determined by $\Delta_{\perp}$ (see Fig.~\ref{fig:conddel}(b)) and barely affected by $\Delta_{\parallel}$ (see Fig.~\ref{fig:conddel}(a)).

\begin{figure}
\includegraphics[width=12cm]{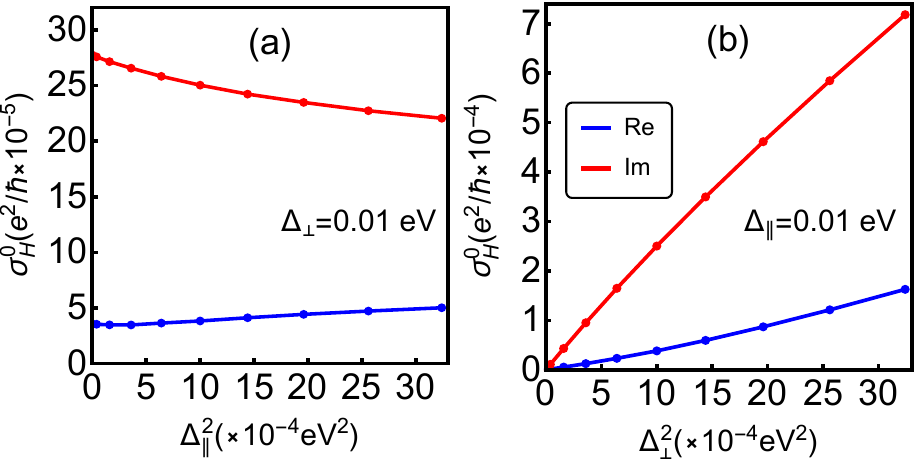}
\caption{The Hall conductivity as a function of (a) $\Delta_{\parallel}^2$ and (b)  $\Delta_{\perp}^2$ for the 3D chiral $p$-wave state. The Hall conductivity is calculated at a fixed frequency $\omega=0.3 $ eV.}
\label{fig:conddel}
\end{figure}

\section{IV. The calculation of the Kerr angle}
To connect our calculations with Kerr rotation experiment in \SRO~\cite{Xia:06}, we evaluate in this section the Kerr angle from the calculated AHC $\sigma_{H}(\omega)$ at $\omega=0.8$ eV. The Kerr angle is given by ~\cite{Taylor:12}
\begin{equation}
\label{Eq:kerr}
\theta_{K}=\frac{4\pi}{\omega d} \text{Im} \left[ \frac{\sigma_{H}}{n(n^2-1)}  \right] \,,
\end{equation}
where $n=n(\omega)$ is the $\omega$-dependent complex index of refraction, given by
\begin{align}
n(\omega) &= \sqrt{\epsilon_{ab}(\omega)}  \\
\epsilon_{ab}(\omega) &= \epsilon_{\infty}+\frac{4\pi i}{\omega} \sigma_{L}(\omega) \,.
\end{align}
Here, $\epsilon_{ab}(\omega)$ is the permeability tensor in the $ab$-plane. $\epsilon_{\infty}=10$ is the background permeability. $d = 6.8 \si{\angstrom}$ is the inter-layer spacing along the $c$-axis. $\sigma_{L}(\omega)$ is the optical longitudinal conductivity. Following Ref.~\cite{Taylor:12} we use a simple Drude model for $\sigma_{L}(\omega)$,
\begin{equation}
\sigma_{L}(\omega)= -\frac{\omega_{pl}^2}{4\pi i(\omega+i\Gamma)} \,,
\end{equation}
where $\omega_{pl} = 2.9$ eV is the the plasma frequency and $\Gamma = 0.4$ eV is an elastic scattering rate. At $ \omega=0.8$ eV, $\sigma_{L}=0.33+i0.67$, $\epsilon_{ab}=0.52+i5.20$, and $n=1.53+i1.69$. Plugging $n$ back into Eq. ~\eqref{Eq:kerr} and using the $\sigma_{H}(\omega=0.8 \text{eV})$ we can obtain $\theta_{K}$.

\section{V. Derivation of Eq.~(13) in the maintext}
In this section, we derive the approximation of Eq.~(13) from Eq.~(12) in the maintext, assuming that only band-1 develops Cooper pairing in our model. In this scenario $|u_{2\bk s}|^2=\theta(\epsilon_{2 \bk})$, $|v_{2\bk s}|^2=\theta(-\epsilon_{2 \bk})$. By making use of $V^{ij}_{\mu\bk s} =(V^{ji}_{\mu\bk s})^\ast$, we obtain
\begin{eqnarray}
&& |u_{1\bk s}|^2|v_{2\bk s}|^2[(V^{12}_{x \bar{\bk} s})^\ast (V^{21}_{y \bar{\bk} s})^\ast-(V^{12}_{y \bar{\bk} s})^\ast (V^{21}_{x \bar{\bk} s})^\ast] +  |u_{2\bk s}|^2|v_{1\bk s}|^2(V^{12}_{x\bk s}V^{21}_{y\bk s}-V^{12}_{y\bk s} V^{21}_{x\bk s})  \nonumber \\ 
&=& \theta(\epsilon_{2\bk})|v_{1\bk s}|^2 (V^{12}_{x\bk s}V^{21}_{y\bk s}-V^{12}_{y\bk s} V^{21}_{x\bk s}) -\theta(-\epsilon_{2\bk})|u_{1\bk s}|^2 (V^{12}_{x\bar{\bk} s}V^{21}_{y\bar{\bk} s}-V^{12}_{y\bar{\bk} s} V^{21}_{x\bar{\bk} s})   \nonumber \\
&=& \theta(\epsilon_{2\bk})|v_{1\bk s}|^2 (V^{12}_{x\bk s}V^{21}_{y\bk s}-V^{12}_{y\bk s} V^{21}_{x\bk s}) - \theta(-\epsilon_{2\bk})(1-|v_{1\bk s}|^2) (V^{12}_{x\bar{\bk} s}V^{21}_{y\bar{\bk} s}-V^{12}_{y\bar{\bk} s} V^{21}_{x\bar{\bk} s})   \nonumber \\
&=& - i\rho_{i\bk s} (\epsilon_{1\bk}-\epsilon_{2\bk})^2[ \theta(\epsilon_{2\bk}) \mathcal{B}^1_{xy,\bk s} + \theta(-\epsilon_{2\bk}) \mathcal{B}^1_{xy,\bar{\bk} s}] +i (\epsilon_{1\bk}-\epsilon_{2\bk})^2\theta(-\epsilon_{2\bk}) \mathcal{B}^1_{xy,\bar{\bk} s} \nonumber \\
& = &- i\rho_{i\bk s} (\epsilon_{1\bk}-\epsilon_{2\bk})^2  \mathcal{B}^1_{xy,\bk s}  \,,  
\label{eq:rhoBerry}
\end{eqnarray}
where $\theta(\epsilon_\bk)$ is the Heaviside step function, and we have used Eq.~(7) in the maintext. Note that, in obtaining the last line, we have dropped the last term in the fourth line, which will be negligibly small if we sum over contributions from different spins to Eq.~(10) in maintext at zero temperature (the two spin contributions exactly cancel if both spin sectors of the nonunitary helical pairing are fully gapped). We have also made use of the inversion and time-reversal symmetries of our normal-state Hamiltonian, which dictate that $\mathcal{B}^i_{xy,\bar{\bk} s } =\mathcal{B}^i_{xy,\bk s} = - \mathcal{B}^i_{xy,\bar{\bk} \bar{s}}$.

\section{VI. Vanishing of anomalous Hall conductivity in unitary chiral superconductors}
In this section, through a derivation of $Q^{ab}_{xy,\bk}$ in Eq. (11) of the maintext for the processes described by Fig. 2, we show how the Hall conductivity vanishes for unitary chiral superconductors without interband pairing. 

A generic BdG Hamiltonian for a two-band unitary chiral superconductor without interband pairing can be written in the Nambu spinor $(c_{1\bk s,},c_{2\bk s,},c^\dagger_{1\bar{\bk} \bar{s}},c^\dagger_{2\bar{\bk} \bar{s}})^T$ as,
\begin{equation}
H_{\bk s} = \begin{pmatrix}
\epsilon_{1\bk}  & 0  & \zeta_s \Delta_{1\bk} & 0 \\
0  & \epsilon_{2\bk} & 0 & \zeta_s \Delta_{2\bk} \\
\zeta_s \Delta_{1\bk}^\ast  & 0 & -\epsilon_{1\bk} & 0 \\
0 & \zeta_s \Delta_{2\bk}^\ast & 0 &-\epsilon_{2\bk} 
\end{pmatrix}
\label{eq:Hunitary}
\end{equation}
Here and after, the notations follow the convention in the maintext, and $\zeta_s =1$ for pseudospin-triplet chiral superconducting pairing, and $\zeta_s = 1~(-1)$ for pseudospin-singlet pairing if the pseudospin index in the Nambu spinor $s=\uparrow~(\downarrow)$. For convenience, we shall later refer to the two Nambu spinors as two sectors. The following derivation will be performed for pseudospin-triplet chiral states. The same analyses can be straightforwardly generalized to pseudospin-singlet states, reaching the same conclusion. 

Consider now the following optical transition processes in the sector-$s$, 
\begin{eqnarray}
&& |\bar{1}\bk s\rangle \xrightarrow{V^{12}_{y\bk}} |2\bk s\rangle  \xrightarrow{V^{21}_{x\bk}} |\bar{1}\bk s\rangle ~~~ - ~~~(x \leftrightarrow y)\,.  \nonumber
\end{eqnarray}
also keeping in mind a subtle point, that is, the corresponding velocity operator must conform with the new spinor basis in \eqref{eq:Hunitary}. Here, the wavefunctions of the quasiparticle states are easily obtained from \eqref{eq:Hunitary}. Specifically, one has 
\begin{equation}
|\bar{1}\bk s\rangle = \begin{pmatrix}
-v^\ast_{1\bk} \\
0\\
u^\ast_{1\bk}\\
0
\end{pmatrix}, ~~~~~~~~|2\bk s\rangle = \begin{pmatrix}
0 \\
u_{2\bk}\\
0\\
v_{2\bk}
\end{pmatrix}
\end{equation}
Note that for triplet pairing the two spinor sectors have the same set of wavefunctions. Following the derivation that led to Eq.~(13) in the maintext, we have for the $Q^{ab}_{xy,\bk}$ generated by these transition processes,
\begin{eqnarray}
+&&u_{2\bk}^\ast v_{1\bk}^\ast u_{1\bk} v_{2\bk} [(V^{12}_{x\bar{\bk} \bar{s}})^\ast V^{21}_{y\bk s}-(V^{12}_{y\bar{\bk} \bar{s}})^\ast V^{21}_{x\bk s}] \nonumber \\
+&& v_{2\bk}^\ast u_{1\bk}^\ast v_{1\bk} u_{2\bk} [V^{12}_{x\bk s} (V^{21}_{y\bar{\bk} \bar{s}})^\ast-V^{12}_{y\bk s} (V^{21}_{x\bar{\bk}\bar{s}})^\ast ] \nonumber \\
+&&|u_{1\bk}|^2|v_{2\bk}|^2[(V^{12}_{x\bar{\bk} \bar{s}})^\ast (V^{21}_{y\bar{\bk} \bar{s}})^\ast-(V^{12}_{y\bar{\bk} \bar{s}})^\ast (V^{21}_{x\bar{\bk} \bar{s}})^\ast ] \nonumber \\
+&&  |u_{2\bk}|^2|v_{1\bk}|^2(V^{12}_{x\bk s}V^{21}_{y\bk s}-V^{12}_{y\bk s} V^{21}_{x\bk s})  \,.
\end{eqnarray}
Making use of the relation $V^{ij}_{\mu\bk s} = -(V^{ij}_{\mu\bar{\bk}\bar{s}})^\ast = (V^{ji}_{\mu\bk s})^\ast$ in our model, the above expression becomes,
\begin{eqnarray}
&&( -u_{2\bk}^\ast v_{1\bk}^\ast u_{1\bk} v_{2\bk}-v_{2\bk}^\ast u_{1\bk}^\ast v_{1\bk} u_{2\bk}+ |u_{1\bk}|^2|v_{2\bk}|^2+|u_{2\bk}|^2|v_{1\bk}|^2) (V^{12}_{x\bk s}V^{21}_{y\bk s}-V^{12}_{y\bk s} V^{21}_{x\bk s})  \\ \nonumber 
&= & -i ( -u_{2\bk}^\ast v_{1\bk}^\ast u_{1\bk} v_{2\bk}-v_{2\bk}^\ast u_{1\bk}^\ast v_{1\bk} u_{2\bk}+ |u_{1\bk}|^2|v_{2\bk}|^2+|u_{2\bk}|^2|v_{1\bk}|^2) (\epsilon_{1\bk}-\epsilon_{2\bk})^2\mathcal{B}^1_{xy,\bk s} \,.
\end{eqnarray}
Since the coefficient in the front is independent of the spinor sector index $s$, and since $\mathcal{B}^1_{xy,\bk s} = -\mathcal{B}^1_{xy,\bk \bar{s}}$ due to time-reversal and inversion symmetries of the normal-state Hamiltonian, the above contribution is exactly opposite to the counterpart processes $|\bar{1}\bk\bar{s}\rangle \rightarrow |2\bk\bar{s}\rangle \rightarrow |\bar{1}\bk \bar{s}\rangle$. We thus see that the quantity $Q^{ab}_{xy,\bk}$, and hence the anomalous Hall conductivity, shall vanish for unitary chiral superconductors.

\end{widetext}

\begin{thebibliography}{99}

\bibitem{Kallin:16} C. Kallin and A. J. Berlinsky. Chiral superconductors. \href{https://doi.org/10.1088/0034-4885/79/5/054502
}{Rep. Prog. Phys. {\bf 79}, 054502 (2016)}.
\bibitem{Xiao:10} D. Xiao, M. C. Chang, and Q. Niu. Berry phase effects on electronic properties.  \href{https://doi.org/10.1103/RevModPhys.82.1959}{Rev. Mod. Phys. 82, 1959 (2010)}.
\bibitem{Nagaosa:10} N. Nagaosa, J. Sinova, S. Onoda, A. H. MacDonald, and N. P. Ong. Anomalous Hall effect.  \href{https://doi.org/10.1103/RevModPhys.82.1539}{Rev. Mod. Phys. 82, 1539 (2010)}.
\bibitem{Xia:06} J. Xia, Y. Maeno, P. T. Beyersdorf, M. M. Fejer, and A. Kapitulnik. High Resolution Polar Kerr Effect Measurements of Sr$_2$RuO$_4$: Evidence for Broken Time-Reversal Symmetry in the Superconducting State. \href{https://doi.org/10.1103/PhysRevLett.97.167002}{Phys. Rev. Lett. {\bf 97}, 167002 (2006)}.
\bibitem{Schemm2014} E. R. Schemm, W. J. Gannon, C. M. Wishne, W. P. Halperin, A. Kapitulnik. Observation of broken time-reversal symmetry in the heavy-fermion superconductor UPt$_3$.  \href{https://doi.org/10.1126/science.1248552}{Science {\bf 345}, 190 (2014)}.
\bibitem{Schemm2015} E. R. Schemm, R. E. Baumbach, P. H. Tobash, F. Ronning, E. D. Bauer, and A. Kapitulnik. Evidence for broken time-reversal symmetry in the superconducting phase of URu$_2$Si$_2$. \href{https://doi.org/10.1103/PhysRevB.91.140506}{Phys. Rev. B 91, 140506 (R) (2015)}.
\bibitem{Read:00} N. Read and D. Green. Paired states of fermions in two dimensions with breaking of parity and time-reversal symmetries and the fractional quantum Hall effect.  \href{https://doi.org/10.1103/PhysRevB.61.10267}{Phys. Rev. {\bf B 61}, 10267 (2000)}.
\bibitem{Karplus:54} R. Karplus and J. M. Luttinger. Hall Effect in Ferromagnetics. \href{https://doi.org/10.1103/PhysRev.95.1154}{Phys. Rev. 95, 1154 (1954)}.
\bibitem{Sundaram:99} G. Sundaram and Q. Niu. Wave-packet dynamics in slowly perturbed crystals: Gradient corrections and Berry-phase effects. \href{https://doi.org/10.1103/PhysRevB.59.14915}{Phys. Rev. B 59, 14915 (1999)}.
\bibitem{Goryo:08} J. Goryo. Impurity-induced polar Kerr effect in a chiral $p$-wave superconductor. \href{https://doi.org/10.1103/PhysRevB.78.060501}{Phys. Rev. B {\bf 78}, 060501(R) (2008)}.
\bibitem{Lutchyn:09} R. M. Lutchyn, P. Nagornykh, and V. M. Yakovenko. Frequency and temperature dependence of the anomalous ac Hall conductivity in a chiral $p_x+ip_y$
 superconductor with impurities. \href{https://doi.org/10.1103/PhysRevB.80.104508}{Phys. Rev. B {\bf 80}, 104508 (2009)}.
\bibitem{Konig:17} E. J. Konig, A. Levchenko. Kerr Effect from Diffractive Skew Scattering in Chiral $p_x\pm ip_y$ Superconductors. \href{https://doi.org/10.1103/PhysRevLett.118.027001}{Phys. Rev. Lett. {\bf 118}, 027001 (2017)}.
\bibitem{LiYu:20} Y. Li, Z. Wang, and W. Huang. Anomalous Hall effect in single-band chiral superconductors from impurity superlattices. \href{https://doi.org/10.1103/PhysRevResearch.2.042027}{Phys. Rev. Res. 2, 042027(R) (2020)}.
\bibitem{Taylor:12} E. Taylor and C. Kallin. Intrinsic Hall Effect in a Multiband Chiral Superconductor in the Absence of an External Magnetic Field. \href{https://doi.org/10.1103/PhysRevLett.108.157001}{Phys. Rev. Lett. {\bf 108}, 157001 (2012)}.
\bibitem{Wysokinski:12} K.I. Wysoki\'nski, J. F. Annett, B. L. Gy\"orffy. Intrinsic Optical Dichroism in the Chiral Superconducting State of Sr$_2$RuO$_4$. \href{https://doi.org/10.1103/PhysRevLett.108.077004}{Phys. Rev. Lett. {\bf 108}, 077004 (2012)}.
\bibitem{Komendova:17} L. Komendova and A. M. Black-Schaffer. Odd-Frequency Superconductivity in Sr$_2$RuO$_4$ Measured by Kerr Rotation.  \href{https://doi.org/10.1103/PhysRevLett.119.087001}{Phys. Rev. Lett. {\bf 119}, 087001 (2017)}.
\bibitem{WangZQ:17} Z. Wang, J. Berlinsky, G. Zwicknagl, and C. Kallin. Intrinsic ac anomalous Hall effect of nonsymmorphic chiral superconductors with an application to UPt$_3$. \href{https://doi.org/10.1103/PhysRevB.96.174511}{Phys. Rev. B {\bf 96}, 174511 (2017)}.
\bibitem{Brydon:19} P. M. R. Brydon, D. S. L. Abergel, D. F. Agterberg, and V. M. Yakovenko. Loop Currents and Anomalous Hall Effect from Time-Reversal Symmetry-Breaking Superconductivity on the Honeycomb Lattice. \href{https://doi.org/10.1103/PhysRevX.9.031025}{Phys. Rev. X {\bf 9}, 031025 (2019)}.
\bibitem{ZhangJL:20} J.L. Zhang, Y. Li, W. Huang, and F. C. Zhang. Hidden anomalous Hall effect in Sr$_2$RuO$_4$ with chiral superconductivity dominated by the Ru $d_{xy}$ orbital. \href{https://doi.org/10.1103/PhysRevB.102.180509}{Phys. Rev. B 102, 180509(R) (2020)}.
\bibitem{Mineev:12} V.P. Mineev. Whether There is the Intrinsic Hall Effect in a Multi-Band Superconductor? \href{https://doi.org/10.1143/JPSJ.81.093703}{J. Phys. Soc. Jpn. 81, 093703 (2012)}.
\bibitem{Denys:21} M. D. E. Denys and P. M. R. Brydon. Origin of the anomalous Hall effect in two-band chiral superconductors. \href{https://doi.org/10.1103/PhysRevB.103.094503}{Phys. Rev. B {\bf 103}, 094503 (2021)}.
\bibitem{Peotta:15} S. Peotta, and P. T\"orm\"a. Superfluidity in topologically nontrivial flat bands. \href{https://doi.org/10.1038/ncomms9944}{Nat. Commun. {\bf 6}, 8944 (2015)}.
\bibitem{Torma:21} P. T\"orm\"a, S. Peotta, and B.A. Bernevig. Superconductivity, superfluidity and quantum geometry in twisted multilayer systems. \href{https://doi.org/10.1038/s42254-022-00466-y
}{Nat. Rev. Phys. \textbf{4}, 528 (2021)}.
\bibitem{AhnJ:22} J. Ahn, G-Y. Guo, N. Nagaosa, and A. Vishwanath. Riemannian geometry of resonant optical responses. \href{https://doi.org/10.1038/s41567-021-01465-z}{Nat. Phys. {\bf 18}, 290 (2022)}.
\bibitem{Liang:17} L. Liang, T. I. Vanhala, S. Peotta, T. Siro, A. Harju, and P. T\"orm\"a. Band geometry, Berry curvature, and superfluid weight. \href{https://doi.org/10.1103/PhysRevB.95.024515}{Phys. Rev. B \textbf{95}, 024515 (2017)}.
\bibitem{Chen:21} W. Chen and W. Huang. Quantum-geometry-induced intrinsic optical anomaly in multiorbital superconductors. \href{https://doi.org/10.1103/PhysRevResearch.3.L042018}{Phys. Rev. Research \textbf{3}, L042018 (2021)}.
\bibitem{Provost:80} J. Provost and G. Vallee. Riemannian structure on manifolds of quantum states. \href{https://doi.org/10.1007/BF02193559}{Commun. Math. Phys. 76, 289 (1980)}.

\bibitem{Maeno2001} Y. Maeno, T. M. Rice, and M. Sigrist. The Intriguing Superconductivity of Strontium Ruthenate. \href{https://doi.org/10.1063/1.1349611}{Phys. Today {\bf 54} (1), 42
(2001)}.

\bibitem{Maeno2003} A. P. Mackenzie and Y. Maeno. The superconductivity of 
Sr$_2$RuO$_4$ and the physics of spin-triplet pairing. \href{https://doi.org/10.1103/RevModPhys.75.657}{Rev. Mod. Phys. {\bf 75}, 657 (2003)}.

\bibitem{Kallin2012} C. Kallin. Chiral p-wave order in Sr$_2$RuO$_4$. \href{http://doi.org/10.1088/0034-4885/75/4/042501}{Rep. Prog. Phys. {\bf 75} 042501 (2012)}.

\bibitem{Maeno2012} Y. Maeno, S. Kittaka, T. Nomura, S. Yonezawa, and K. Ishida. Evaluation of Spin-Triplet Superconductivity in Sr$_2$RuO$_4$. \href{https://doi.org/10.1143/JPSJ.81.011009}{J. Phys. Soc. Jpn. {\bf 81}, 011009 (2012)}.

\bibitem{Mac2017} A. P. Mackenzie, T. Scaffidi, C. W. Hicks and Y. Maeno. Even odder after twenty-three years: the superconducting order parameter puzzle of Sr$_2$RuO$_4$.  \href{https://doi.org/10.1038/s41535-017-0045-4}{npj Quant Mater {\bf 2}, 40 (2017)}.

\bibitem{Huang2021CPB} W. Huang. A review of some new perspectives on the theory of superconducting Sr$_2$RuO$_4$. \href{https://doi.org/10.1088/1674-1056/ac2488}{Chin. Phys. B, {\bf 30}, 107403 (2021)}.

\bibitem{Leggett2021} A. J. Leggett, Y. Liu. Symmetry Properties of Superconducting Order Parameter in Sr$_2$RuO$_4$.  \href{https://doi.org/10.1007/s10948-020-05717-6}{J Supercond Nov Magn {\bf 34}, 1647–1673 (2021)}. 

\bibitem{Luke1998} G. M. Luke, Y. Fudamoto, K. M. Kojima, M. I. Larkin, J. Merrin, B. Nachumi, Y. J. Uemura, Y. Maeno, Z. Q. Mao, Y. Mori, H. Nakamura, and M. Sigrist. Time-reversal symmetry-breaking superconductivity in Sr$_2$RuO$_4$.  \href{https://doi.org/10.1038/29038}{Nature {\bf 394}, 558–561 (1998)}.

\bibitem{Ishida1998} K. Ishida, H. Mukuda, Y. Kitaoka, K. Asayama, Z. Q. Mao, Y. Mori and Y. Maeno. Spin-triplet superconductivity in Sr$_2$RuO$_4$ identified by $^{17}$O Knight shift. \href{https://doi.org/10.1038/25315}{Nature {\bf 396}, 658–660 (1998)}.

\bibitem{Duffy2000} J. A. Duffy, S. M. Hayden, Y. Maeno, Z. Mao, J. Kulda, and G. J. McIntyre. Polarized-Neutron Scattering Study of the Cooper-Pair Moment in Sr$_2$RuO$_4$. \href{https://doi.org/10.1103/PhysRevLett.85.5412}{Phys. Rev. Lett. {\bf 85}, 5412 (2000)}.

\bibitem{Liu2004} K. D. Nelson, Z. Q. Mao, Y. Maeno, Y. Liu. Odd-Parity Superconductivity in Sr$_2$RuO$_4$. \href{https://doi.org/10.1126/science.1103881}{Science {\bf 306}, 1151-1154(2004)}.

\bibitem{Hicks:14} C.W. Hicks, D.O. Brodsky, E.A. Yelland, A.S. Gibbs, J.A.N. Bruin, M.E. Barber, S.D. Edkins, K. Nishimura, S. Yonezawa, Y. Maeno, A.P. Mackenzie, Strong Increase of $T_c$ of \SRO~Under Both Tensile and Compressive Strain. \href{https://www.science.org/doi/10.1126/science.1248292}{Science {\bf 344}, 283 (2014)}.

\bibitem{Pus2019} A. Pustogow, Yongkang Luo, A. Chronister, Y.-S. Su, D. A. Sokolov, F. Jerzembeck, A. P. Mackenzie, C. W. Hicks, N. Kikugawa, S. Raghu, E. D. Bauer and S. E. Brown. Constraints on the superconducting order parameter in Sr$_2$RuO$_4$ from oxygen-17 nuclear magnetic resonance. \href{https://doi.org/10.1038/s41586-019-1596-2}{Nature {\bf 574}, 72–75 (2019)}.

\bibitem{Chronister2021} A. Chronister, A. Pustogow, N. Kikugawa, D. A. Sokolov, F. Jerzembeck, C. W. Hicks, A. P. Mackenzie, E. D. Bauer, and S. E. Brown. Evidence for even parity unconventional superconductivity in Sr$_2$RuO$_4$.  \href{https://doi.org/10.1073/pnas.2025313118}{Proc. Natl. Acad. Sci. (USA) {\bf 118}, e2025313118 (2021)}.

\bibitem{exp1} E. Hassinger, P. Bourgeois-Hope, H. Taniguchi, S. Ren\'{e} de Cotret, G. Grissonnanche, M. S. Anwar, Y. Maeno, N. Doiron-Leyraud, and L. Taillefer. Vertical Line Nodes in the Superconducting Gap Structure of Sr$_2$RuO$_4$. \href{https://doi.org/10.1103/PhysRevX.7.011032}{Phys. Rev. X {\bf 7}, 011032 (2017)}.

\bibitem{exp3} Y. S. Li, M. Garst, J. Schmalian, S. Ghosh, N. Kikugawa, D. A. Sokolov, C. W. Hicks, F. Jerzembeck, M. S. Ikeda, Z. Hu, B. J. Ramshaw, A. W. Rost, M. Nicklas and A. P. Mackenzie. Elastocaloric determination of the phase diagram of Sr$_2$RuO$_4$. \href{https://doi.org/10.1038/s41586-022-04820-z}{Nature {\bf 607}, 276 (2022)}.

\bibitem{ultrasound1}  S. Benhabib, C. Lupien, I. Paul, L. Berges, M. Dion, M. Nardone, A. Zitouni, Z. Q. Mao, Y. Maeno, A. Georges, L. Taillefer and C. Proust. Ultrasound evidence for a two-component superconducting order parameter in Sr$_2$RuO$_4$. \href{https://doi.org/10.1038/s41567-020-1033-3}{Nature Physics {\bf 17}, 194 (2021)}.

\bibitem{ultrasound2}  S. Ghosh, A. Shekhter, F. Jerzembeck, N. Kikugawa, D. A. Sokolov, M. Brando, A. P. Mackenzie, C. W. Hicks and B. J. Ramshaw. Thermodynamic evidence for a two-component superconducting order parameter in Sr$_2$RuO$_4$. \href{https://doi.org/10.1038/s41567-020-1032-4}{Nature Physics {\bf 17}, 199 (2021)}.
\bibitem{Cho:16} W. Cho and S.A. Kivelson. Necessity of Time-Reversal Symmetry Breaking for the Polar Kerr Effect in Linear Response. \href{https://doi.org/10.1103/PhysRevLett.116.093903}{Phys. Rev. Lett. {\bf 116}, 093903 (2016)}.
\bibitem{Huang2021prr} W. Huang and Z. Wang. Possibility of mixed helical p-wave pairings in Sr$_2$RuO$_4$. \href{https://doi.org/10.1103/PhysRevResearch.3.L042002}{Phys. Rev. Res. {\bf 3}, L042002 (2021)}.
\bibitem{footnote} Superconducting states such as $s+id_{x^2-y^2}$ and $d_{x^2-y^2}+ig_{xy(x^2-y^2)}$ in pristine samples of \SRO~cannot generate Kerr effect, as they each preserve certain vertical mirror plane symmetry. This holds true even in the presence of random disorder~\cite{LiuHT:23}. The Kerr effect may emerge if the sample contains special type of defects that break the remaining vertical mirror plane symmetry~\cite{Kivelson:20}. However, the defect must be sufficiently dense to reconcile with the observed magnitude of Kerr rotation -- an unlikely scenario given the high sensitivity of superconductivity to disorder. 
\bibitem{LiuHT:23} H-T. Liu, W. Chen, and W. Huang. Impact of random impurities on the anomalous Hall effect in chiral superconductors. \href{https://doi.org/10.1103/PhysRevB.107.224517}{Phys. Rev. B {\bf 107}, 224517 (2023)}. 
\bibitem{Kivelson:20} S. A. Kivelson, A. C. Yuan, B. Ramshaw, and R. Thomale. A proposal for reconciling diverse experiments on the superconducting state in Sr$_2$RuO$_4$. \href{https://doi.org/10.1038/s41535-020-0245-1}{npj Quantum Materials {\bf 5}, 43 (2020)}.
\bibitem{Huang2018prl} W. Huang and H. Yao. Possible Three-Dimensional Nematic Odd-Parity Superconductivity in Sr$_2$RuO$_4$. \href{https://doi.org/10.1103/PhysRevLett.121.157002}{Phys. Rev. Lett. {\bf 121}, 157002 (2018)}. 
\bibitem{footnote2} Note that Fig.~\ref{fig:Berry} was obtained using a constant SOC $\eta_\bk s_z \equiv \eta_0 s_z$. In principle, this term permits higher-order harmonics such as $\eta_1(\cos k_x + \cos k_y)s_z$. With finite $\eta_1$, the Hamiltonian Eq.~\eqref{eq:2bandTB} resembles that of a quantum spin Hall insulating state with spin Chern number 0, $\pm 2$ for the individual bands. 
\bibitem{Imai:12} Y. Imai, K. Wakabayashi, and M. Sigrist. Properties of edge states in a spin-triplet two-band superconductor. \href{https://doi.org/10.1103/PhysRevB.85.174532}{Phys. Rev. B {\bf 85}, 174532 (2012)}.
\bibitem{Haldane:04} F.D.M. Haldane. Berry Curvature on the Fermi Surface: Anomalous Hall Effect as a Topological Fermi-Liquid Property. \href{https://doi.org/10.1103/PhysRevLett.93.206602}{Phys. Rev. Lett. 93, 206602 (2004)}.  
\bibitem{supplementary} See Supplementary Materials for more details of the model formulation. 
\bibitem{Sharma:20} R. Sharma, S. Edkins, Z. Wang, A. Kostin, C. Sow, Y. Maeno,
A. Mackenzie, J. C. Seamus Davis, and V. Madhavan. Momentum-resolved superconducting energy gaps of Sr$_2$RuO$_4$ from quasiparticle interference imaging. \href{https://doi.org/10.1073/pnas.1916463117}{Proc. Natl. Acad. Sci. USA {\bf 117}, 5222 (2020)}.
\bibitem{Hsu2011} C.-H. Hsu, S. Raghu, and S. Chakravarty, Topological density wave states of nonzero angular momentum. \href{https://journals.aps.org/prb/abstract/10.1103/PhysRevB.84.155111}{Phys. Rev. B 84, 155111 (2011)}
\end{thebibliography}
\end{document}